\documentclass[12pt]{cernrep}
\usepackage{epsf,epsfig,cite}

\newcount\timecount
\newcount\hours \newcount\minutes  \newcount\temp \newcount\pmhours

\input paperdef
\def\citere{\cite}
\def\citeres{\cite}

\begin{document}

\thispagestyle{empty}
\setcounter{page}{0}
\def\thefootnote{\fnsymbol{footnote}}

\begin{flushright}
CERN--PH--TH/2006--068\\
DCPT/06/48, IPPP/06/24\\
FTPI--MINN--06/12, UMN--TH--2440/06\\ 
hep-ph/0604180\\
\end{flushright}

\vspace{1cm}

\begin{center}

{\large\sc {\bf Indications of the CMSSM Mass Scale from Precision
    Electroweak Data}}%
\footnote{Contribution to the {\it 2005 International Linear Collider
    and Detector Workshop}, Snowmass, Colorado,\\ 
 \mbox{}\hspace{0.7cm}August 14-27, 2005}

\vspace{1cm}

{\sc J.~Ellis$^{1\,}$%
\footnote{
email: John.Ellis@cern.ch 
}%
, S.~Heinemeyer$^{\,2}$%
\footnote{
email: Sven.Heinemeyer@cern.ch
}%
, K.A.~Olive$^{\,3}$%
\footnote{
email: olive@mnhep.hep.umn.edu
}%
~and G.~Weiglein$^{\,4}$%
\footnote{
email: Georg.Weiglein@durham.ac.uk
}%
}

\vspace*{1cm}

$^1$ TH Division, Physics Department, CERN, Geneva, Switzerland

\vspace*{0.4cm}

$^2$ Depto.\ de F\'isica Te\'orica, Universidad de Zaragoza, 50009 Zaragoza,
Spain 

\vspace*{0.4cm}

$^3$ William I.\ Fine Theoretical Physics Institute,\\
University of Minnesota, Minneapolis, MN~55455, USA

\vspace*{0.4cm}

$^4$ IPPP, University of Durham, Durham DH1~3LE, UK

\end{center}

\vspace*{1cm}

\begin{abstract}
We discuss the sensitivities of present-day electroweak precision data to
the possible scale of supersymmetry within
the constrained minimal supersymmetric extension of the Standard Model (CMSSM).
Our analysis is based on $\MW$, 
$\sweff$, $(g-2)_\mu$, $\br(b \to s \ga)$, and the lightest
MSSM Higgs boson mass, $\Mh$. We display the impact of the recent
reduction in $\mt$ from $178.0 \pm 4.3 \gev$ to $172.7 \pm 2.9 \gev$
on the interpretation of the precision observables. We show the
currently preferred values of the CMSSM mass scale  $m_{1/2}$ based on
a global $\chi^2$ fit, assuming that the lightest supersymmetric
particle (LSP) is a neutralino, and fixing $m_0$ so as to obtain the
cold dark matter density allowed by WMAP and other cosmological data
for specific values of $A_0$, $\tb$ and $\mu > 0$. The recent
reduction in $\mt$ reinforces previous indications for relatively
light soft supersymmetry-breaking masses, offering good prospects for
the LHC and the ILC, and in some cases also for the Tevatron. Finally,
we discuss the sensitivity of the global $\chi^2$ function to
possible future evolution in the experimental central value of $\mt$
and its error.  
\end{abstract}

\def\thefootnote{\arabic{footnote}}
\setcounter{footnote}{0}

%%%%%%%%%%%%%%%%%%%%%%%%%%%%%%%%%%%%%%%%%%%%%%%%%%%%%%%%%%%%%%%%%%%%%
%%%%%%%%%%%%%%%%%%%%%%%%%%%%%%%%%%%%%%%%%%%%%%%%%%%%%%%%%%%%%%%%%%%%%

\newpage

\title{Indications of the CMSSM Mass Scale from Precision Electroweak Data}

\author{J.~Ellis$^1$, S.~Heinemeyer$^2$, K.A.~Olive$^3$, G.~Weiglein$^4$}
\institute{
$^1${TH Division, Physics Department, CERN, Geneva, Switzerland}\\
$^2$Depto.\ de F\'isica Te\'orica, Universidad de Zaragoza, 50009 Zaragoza,
Spain \\
$^3${William I.\ Fine Theoretical Physics Institute,
University of Minnesota, Minneapolis, \mbox{}~MN~55455, USA}\\
$^4${IPPP, University of Durham, Durham DH1~3LE, UK}\\
}
\maketitle

\begin{abstract}
We discuss the sensitivities of present-day electroweak precision data to
the possible scale of supersymmetry within
the constrained minimal supersymmetric extension of the Standard Model (CMSSM).
Our analysis is based on $\MW$, 
$\sweff$, $(g-2)_\mu$, $\br(b \to s \ga)$, and the lightest
MSSM Higgs boson mass, $\Mh$. We display the impact of the recent
reduction in $\mt$ from $178.0 \pm 4.3 \gev$ to $172.7 \pm 2.9 \gev$
on the interpretation of the precision observables. We show the
currently preferred values of the CMSSM mass scale  $m_{1/2}$ based on
a global $\chi^2$ fit, assuming that the lightest supersymmetric
particle (LSP) is a neutralino, and fixing $m_0$ so as to obtain the
cold dark matter density allowed by WMAP and other cosmological data
for specific values of $A_0$, $\tb$ and $\mu > 0$. The recent
reduction in $\mt$ reinforces previous indications for relatively
light soft supersymmetry-breaking masses, offering good prospects for
the LHC and the ILC, and in some cases also for the Tevatron. Finally,
we discuss the sensitivity of the global $\chi^2$ function to
possible future evolution in the experimental central value of $\mt$
and its error.  
\end{abstract}

%%%%%%%%%%%%%%%%%%%%%%%%%%%%%%%%%%%%%%%%%%%%%%%%%%%%%%%%%%%%%%%%%%%%%
%%%%%%%%%%%%%%%%%%%%%%%%%%%%%%%%%%%%%%%%%%%%%%%%%%%%%%%%%%%%%%%%%%%%%

\section{Introduction}

We have recently analyzed the indications provided by current experimental
data concerning the possible scale of
supersymmetry~\cite{ehow3,LCWS05ehow3,ehow4} within the framework of
the minimal supersymmetric extension of the Standard Model
(MSSM)~\cite{susy,susy2}. We focus on the constrained MSSM
(CMSSM), in which it is assumed that the soft  
supersymmetry-breaking scalar masses $m_0$, gaugino
masses $m_{1/2}$ and tri-linear parameters $A_0$ are each constrained to
be universal at the input GUT scale, with the gravitino heavy and the
lightest supersymmetric particle (LSP) being the lightest neutralino
$\neu{1}$. 

It is well known that predicting the masses of supersymmetric particles
using precision low-energy data is more difficult than it was for the top
quark or even the Higgs boson. This is because the Standard Model (SM) is
renormalizable, so decoupling theorems imply that many low-energy
observables are insensitive to heavy sparticles~\cite{decoupling}. On
the other hand, 
supersymmetry may provide an important contribution to loop-induced
processes. In fact, it was
found~\cite{ehow3,ehow4} that present data on the electroweak
precision 
observables $\MW$ and $\sweff$, as well as the loop-induced quantities
$(g-2)_\mu$ and $\br(b \to s \ga)$ (see \citere{PomssmRep} for a review),
may already be providing interesting indirect information on the scale of
supersymmetry breaking, at least within the context of the CMSSM with a
neutralino LSP. In that framework, the range of $m_0$ is very restricted
by the cold dark matter density $\Omega_\chi h^2$ determined by WMAP and
other observations, for any set of assumed values of $\tb, m_{1/2}$ and
the trilinear soft supersymmetry-breaking parameter
$A_0$~\cite{WMAPstrips,them}: in our analysis we have fixed $m_0$ 
to satisfy the cold dark matter density constraint, 
$0.094 < \Omega_{\rm CDM} h^2 < 0.129$~\cite{WMAP}~%
\footnote{
The central value of 
$\Omega_{\rm CDM} h^2$ indicated  by the recent three-year WMAP data
is very similar, whilst the uncertainty is now somewhat
reduced~\cite{WMAP3}.
}%
. 

Within the CMSSM and using the (then) preferred range $\mt = 178.0 \pm 4.3
\gev$~\cite{oldmt}, we found previously~\cite{ehow3,LCWS05ehow3} a
preference for low values of $m_{1/2}$, particularly for $\tb = 10$, that
exhibited only a moderate sensitivity to $A_0$~
\footnote{
Our notation for the $A_0$ parameter follows that
which is standard in supergravity models (see e.g.\ \citere{susy}),
namely the coupling in the scalar potential is given by
$A_0 \, g^{(3)}$ for the tri-linear
superpotential term $g^{(3)}$. This differs from the sign
convention used in many publicly available codes, see e.g.\ \citere{spa}. 
}.%
~Here we focus on the change induced by the decrease of the
experimental value of $\mt$. The new analysis~\cite{ehow4} updates our
previous analysis~\cite{ehow3}, taking into account the experimental
result of $\mt = 172.7 \pm 2.9 \gev$~\cite{newmt}, and
provides a {\it vade mecum} for understanding the implications of any
further evolution in the preferred range and experimental error
of $\mt$~%
\footnote{
We also briefly comment on the effect of using the most up-to-date
value of $\mt = 172.5 \pm 2.3 \gev$~\cite{newestmt}.
}%
. 

As we show here explicitly, the new experimental value of $\mt$ has a
non-trivial effect on the ranges of $m_{1/2}$ preferred by the
experimental measurements of $\MW$ and $\sweff$. 
Moreover, it reduces substantially the mass expected for the lightest MSSM
Higgs boson, $\Mh$, for any given values of $m_{1/2}, m_0, \tb$ and $A_0$,
thereby strengthening the constraints on $m_{1/2}$. 
We therefore improve our analysis by incorporating
the full likelihood information provided by the final
results of the LEP search for a Standard Model-like Higgs
boson~\cite{LEPHiggsSM,LEPHiggsMSSM} (see \citere{other} for other recent
analyses in the framework of the CMSSM, which differ
from our analysis by the treatment of certain observables such as $\MW$,
$\sweff$ or $\Mh$, or in their treatment of the 95\% C.L.\ exclusion
bound for $\Mh$.)

%%%%%%%%%%%%%%%%%%%%%%%%%%%%%%%%%%%%%%%%%%%%%%%%%%%%%%%%%%%%%%%%%%%%%
%%%%%%%%%%%%%%%%%%%%%%%%%%%%%%%%%%%%%%%%%%%%%%%%%%%%%%%%%%%%%%%%%%%%%

\section{Current experimental data}

In this Section we review briefly the experimental
data set that has been used for the fits. We focus on parameter points that
yield the correct value of the cold dark matter density, 
$0.094 < \Omega_{\rm CDM} h^2 < 0.129$~\cite{WMAP}, which is, however, not
included in the fit itself. 
The data set furthermore comprises the following
observables: the mass of the $W$~boson, $\MW$, the effective leptonic
weak mixing angle, $\sweff$, the anomalous magnetic moment of the
muon, $(g-2)_\mu$, the radiative $B$-decay
branching ratio $\br(b \to s \ga)$, and the lightest MSSM Higgs boson
mass, $\Mh$.
A detailed description of the first four observables can be found
in~\cite{ehow3,PomssmRep}.  
We limit ourselves here to recalling the current precision of
the experimental results and the theoretical predictions. 
The experimental values of these observables have not changed
significantly compared to~\cite{ehow3,PomssmRep}, 
and neither have the theoretical calculations. However, the lower
experimental value for $\mt$ affects the interpretation of $\MW$ and
$\sweff$, in particular, changing the room available for contributions
from possible physics beyond the Standard Model, such as
supersymmetry. Moreover, as already commented, the new, lower
experimental value of $\mt$ necessitates the incorporation of more
complete experimental information about $\Mh$ into the  
fit. 

The uncertainties in the precision observables are given as follows:

\begin{itemize}

%%%%%%%%%%%%%%%%%%%%%%%%%%%%%%%%%%%%%%%%%%%%%%%%%%%%%%%%%%%%%%%%%%%%%
%%%%%%%%%%%%%%%%%%%%%%%%%%%%%%%%%%%%%%%%%%%%%%%%%%%%%%%%%%%%%%%%%%%%%

\item {\it The $W$~boson mass}:\\
The intrinsic theoretical uncertainty in the prediction for
$\MW$ within the MSSM with real parameters has been estimated
as~\cite{drMSSMal2} 
\BE
\De\MW^{\rm intr,current} \lsim 9 \mev~,
\EE
depending on the mass scale of the supersymmetric particles. A recent
reevaluation of $\MW$~\cite{MWweber}, taking into account all existing
corrections 
yields results very similar (within $\sim 5 \mev$) to our calculation.
The parametric uncertainties are dominated by the experimental error of
the top-quark mass
and the hadronic contribution to the shift in the
fine structure constant. Their current errors induce the following
parametric uncertainties~\cite{lcwsSLACmt,PomssmRep}
\BEA
\de\mt^{\rm current} = 2.9 \; (2.3) \gev &\Rightarrow&
\De\MW^{{\rm para},\mt, {\rm current}} \approx 17.5 \; (14) \mev,  \\[.3em]
\de(\De\al_{\rm had}^{\rm current}) = 36 \times 10^{-5} &\Rightarrow&
\De\MW^{{\rm para},\De\al_{\rm had}, {\rm current}} \approx 6.5 \mev~.
\EEA
The experimental value of $\MW$ used in this analysis 
is~\cite{lepewwg,LEPEWWG}~%
\footnote{
The newest experimental value of 
$\MW^{\rm exp} = 80.404 \pm 0.030 \gev$~\cite{LEPEWWG}
yields practically identical results.
}%
\BE
\MW^{\rm exp,current} = 80.410 \pm 0.032 \gev.
\label{mwexp}
\EE
The experimental and theoretical errors for $\MW$ 
are added in quadrature in our analysis.

%%%%%%%%%%%%%%%%%%%%%%%%%%%%%%%%%%%%%%%%%%%%%%%%%%%%%%%%%%%%%%%%%%%%%
%%%%%%%%%%%%%%%%%%%%%%%%%%%%%%%%%%%%%%%%%%%%%%%%%%%%%%%%%%%%%%%%%%%%%

\item {\it The effective leptonic weak mixing angle}:\\
In the MSSM, the remaining intrinsic theoretical uncertainty in the
prediction  for $\sweff$ has been estimated as~\cite{drMSSMal2}
\BE
\De\sweff^{\rm intr,current} \lsim 7 \times 10^{-5}, 
\EE
depending on the supersymmetry mass scale.
The current experimental errors of $\mt$ and $\De\al_{\rm had}$
induce the following parametric uncertainties
\BEA
\de\mt^{\rm current} = 2.9 \; (2.3) \gev &\Rightarrow&
\De\sweff^{{\rm para},\mt, {\rm current}} 
                           \approx 10 \; (8) \times 10^{-5},  \\[.3em]
\de(\De\al_{\rm had}^{\rm current}) = 36 \times 10^{-5} &\Rightarrow&
\De\sweff^{{\rm para},\De\al_{\rm had}, {\rm current}} \approx 
13 \times 10^{-5} .
\EEA
The experimental value is~\cite{lepewwg,LEPEWWG}
\BE
\sweff^{\rm exp,current} = 0.23153 \pm 0.00016~.
\label{swfit}
\EE
The experimental and theoretical errors for $\sweff$ 
are added in quadrature in our analysis.

%%%%%%%%%%%%%%%%%%%%%%%%%%%%%%%%%%%%%%%%%%%%%%%%%%%%%%%%%%%%%%%%%%%%%
%%%%%%%%%%%%%%%%%%%%%%%%%%%%%%%%%%%%%%%%%%%%%%%%%%%%%%%%%%%%%%%%%%%%%

\item {\it The anomalous magnetic moment of the muon}:\\
We use here the latest
estimate based on $e^+e^-$ data~\cite{Hocker:2004xc} (see
\citeres{g-2review,g-2review2} for reviews):
\BE
\amutheo = 
(11\, 659\, 182.8 \pm 6.3_{\rm had} \pm 3.5_{\rm LBL} \pm 0.3_{\rm QED+EW})
 \times 10^{-10},
\label{eq:amutheo}
\EE
where the source of each error is labelled.

The result for the SM prediction is to be compared with
the final result of the Brookhaven $(g-2)_\mu$ experiment 
E821~\cite{g-2exp,g-2exp2}, namely:
\BE
\amuexp = (11\, 659\, 208.0 \pm 5.8) \times 10^{-10},
\label{eq:amuexp}
\EE
leading to an estimated discrepancy
\BE
\amuexp-\amutheo = (25.2 \pm 9.2) \times 10^{-10},
\label{delamu}
\EE
equivalent to a 2.7~$\sigma$ effect. 
While it would be
premature to regard this deviation as a firm evidence for new
physics,
it does indicate a preference for a non-zero supersymmetric contribution.

We note that new $e^+e^-$ data sets
have recently been published in~\cite{KLOE,CMD2,SND,SND2}, but not yet used in
an updated estimate of $(g - 2)_\mu$. Their inclusion is not expected to alter
substantially the estimate given in (\ref{eq:amutheo}). In particular,
we note that the SND data~\cite{SND} have recently been revised
significantly~\cite{SND2}, following a re-evaluation of the background
processes $e^+ e^- \to \pi^+ \pi^- \gamma$ and $\mu^+ \mu^-
\gamma$. They are now in much better agreement with the CMD2
data~\cite{CMD2}, and show an increased disagreement with the $\tau$
decay data~%
\footnote{
We thank Lee Roberts for information on this point.
}%
. 

%%%%%%%%%%%%%%%%%%%%%%%%%%%%%%%%%%%%%%%%%%%%%%%%%%%%%%%%%%%%%%%%%%%%%
%%%%%%%%%%%%%%%%%%%%%%%%%%%%%%%%%%%%%%%%%%%%%%%%%%%%%%%%%%%%%%%%%%%%%

\item {\it The decay $b \to s \ga$}:\\
Since this decay occurs at the loop level in the SM, the MSSM 
contribution might {\it a priori} be of similar magnitude. A
recent
theoretical estimate of the SM contribution to the branching ratio
is~\cite{hugr}
\BE
\br( b \to s \ga ) = (3.70 \pm 0.46) \times 10^{-4},
\label{bsga}
\EE
where the calculations have been carried out completely to NLO in the 
\msbar\ renormalization scheme~\cite{ali,bsgKO2,ali2}, and the error is
dominated by  
higher-order QCD uncertainties. We record, however, that the error 
estimate for $\br(b \to s \ga)$ is still under theoretical debate, see
also \citeres{hulupo,bsgneubert}. 

For the experimental value, we assume~\cite{ehow4} the estimate~\cite{bsgexp}
\BE
\br( b \to s \ga ) = (3.39^{+ 0.30}_{- 0.027}) \times 10^{-4},
\label{bsgaexp}
\EE
whereas the present experimental 
value estimated just recently by the Heavy Flavour Averaging Group (HFAG)
is $\br( b \to s \ga ) = (3.55 \pm 0.24^{+ 0.09}_{-0.10} \pm 0.03)
\times 10^{-4}$~\cite{HFAG}. The uncertainties are combined
statistical and systematic errors, the systematic error due to the
spectral shape function,  and the uncertainty due to the $d \gamma$
fraction, respectively. The new central value is somewhat closer to
that in the SM (\ref{bsga}), imposing a somewhat stronger 
constraint on the supersymmetric mass scale, but we do not expect the
conclusion to differ greatly from this analysis.

%%%%%%%%%%%%%%%%%%%%%%%%%%%%%%%%%%%%%%%%%%%%%%%%%%%%%%%%%%%%%%%%%%%%%
%%%%%%%%%%%%%%%%%%%%%%%%%%%%%%%%%%%%%%%%%%%%%%%%%%%%%%%%%%%%%%%%%%%%%

\item {\it The lightest MSSM Higgs boson mass}:\\
The mass of the lightest $\cp$-even MSSM Higgs boson can be predicted in 
terms of
the other CMSSM parameters. At the tree level, the two $\cp$-even Higgs 
boson masses are obtained as functions of $\MZ$, the $\cp$-odd Higgs
boson mass $\MA$, and $\tb$. 
For the theoretical prediction of $\Mh$ we employ the
Feynman-diagrammatic method using the code 
{\tt FeynHiggs}~\cite{feynhiggs,feynhiggs2},
which includes all numerically relevant known higher-order corrections.
The current intrinsic error of $\Mh$ due to
unknown higher-order corrections has been estimated to 
be~\cite{mhiggsAEC,mhiggsFDalbals,PomssmRep}
\BE
\De\Mh^{\rm intr,current} = 3 \gev~.
\EE
Details about the
inclusion of $\Mh$ and the evaluation of the corresponding $\chi^2$
values obtained from the direct searches for a Standard Model (SM) Higgs 
boson at LEP~\cite{LEPHiggsSM} can be found in \citere{ehow4}.

\end{itemize}

Assuming that the five observables listed above are
uncorrelated, a $\chi^2$ fit has been performed with
\BE
\chi^2 \equiv \sum_{n=1}^{4} \KL \frac{R_n^{\rm exp} - R_n^{\rm theo}}
                                 {\si_n} \KR^2 + \chi^2_{\Mh}.
\label{eq:chi2}
\EE
Here $R_n^{\rm exp}$ denotes the experimental central value of the
$n$th observable ($\MW$, $\sweff$, \mbox{$(g-2)_\mu$} and $\br(b \to s \ga)$),
$R_n^{\rm theo}$ is the corresponding CMSSM prediction and $\si_n$
denotes the combined error, and $\chi^2_{\Mh}$ denotes the
$\chi^2$ contribution coming from the lightest MSSM Higgs boson
mass~\cite{ehow4}. 

%%%%%%%%%%%%%%%%%%%%%%%%%%%%%%%%%%%%%%%%%%%%%%%%%%%%%%%%%%%%%%%%%%%%%
%%%%%%%%%%%%%%%%%%%%%%%%%%%%%%%%%%%%%%%%%%%%%%%%%%%%%%%%%%%%%%%%%%%%%

\section{CMSSM analysis for {\boldmath $\mt = 172.7 \gev$}}
\label{sec:CMSSMupdate}

As already mentioned, in our old analysis of the CMSSM~\cite{ehow3} 
we used the range $\mt = 178.0 \pm 4.3 \gev$ that was then preferred
by direct measurements~\cite{oldmt}. 
The preferred range has subsequently evolved to 
$172.7 \pm 2.9 \gev$~\cite{newmt} (and very recently to 
$172.5 \pm 2.3 \gev$~\cite{newestmt}). 
The effect of this lower $\mt$ value is twofold. 

\begin{figure}[htb!]
\begin{center}
%\vspace{-0.5cm}
\includegraphics[width=.45\textwidth,height=6cm]{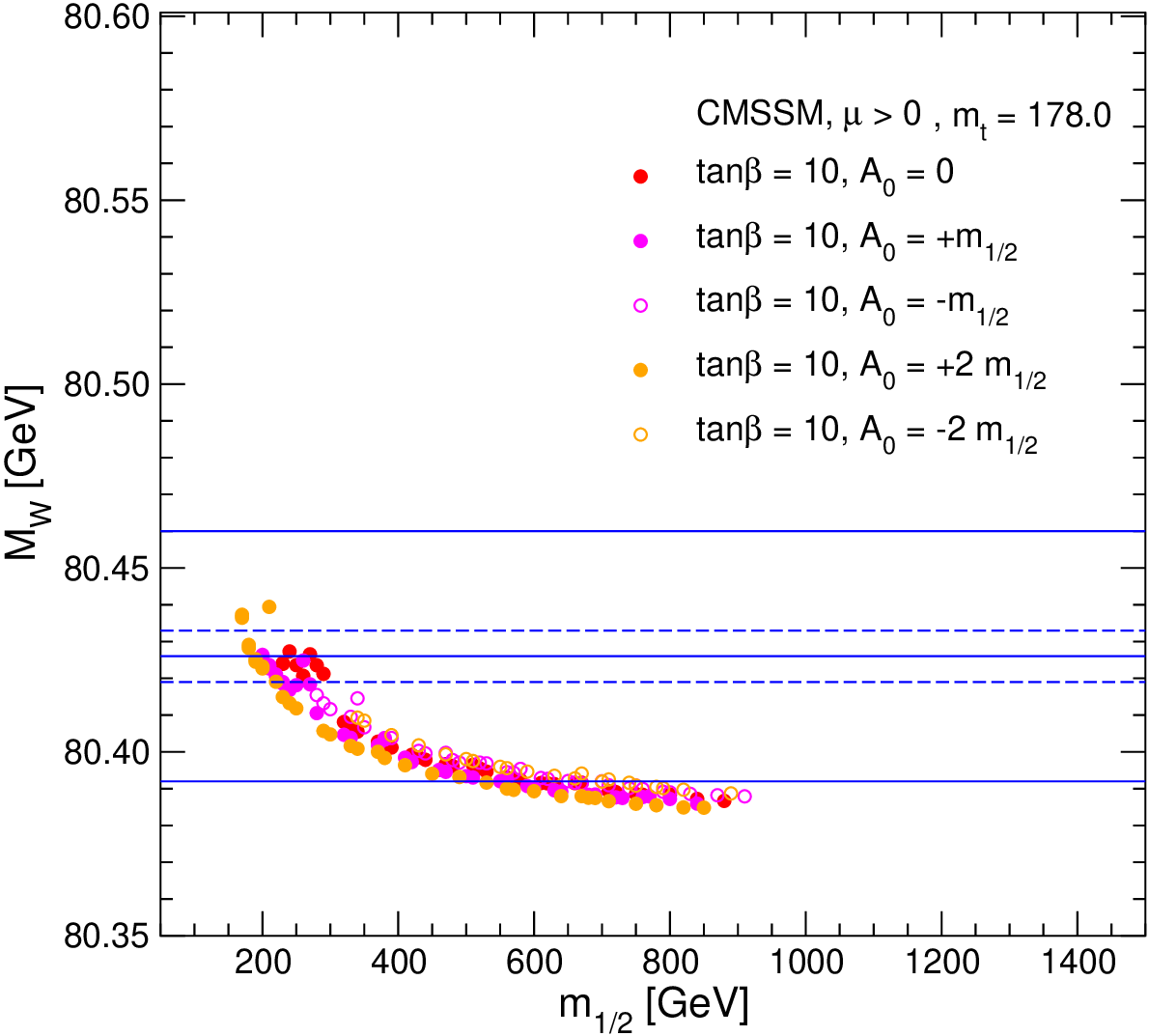}\hspace{1em}
\includegraphics[width=.45\textwidth,height=6cm]{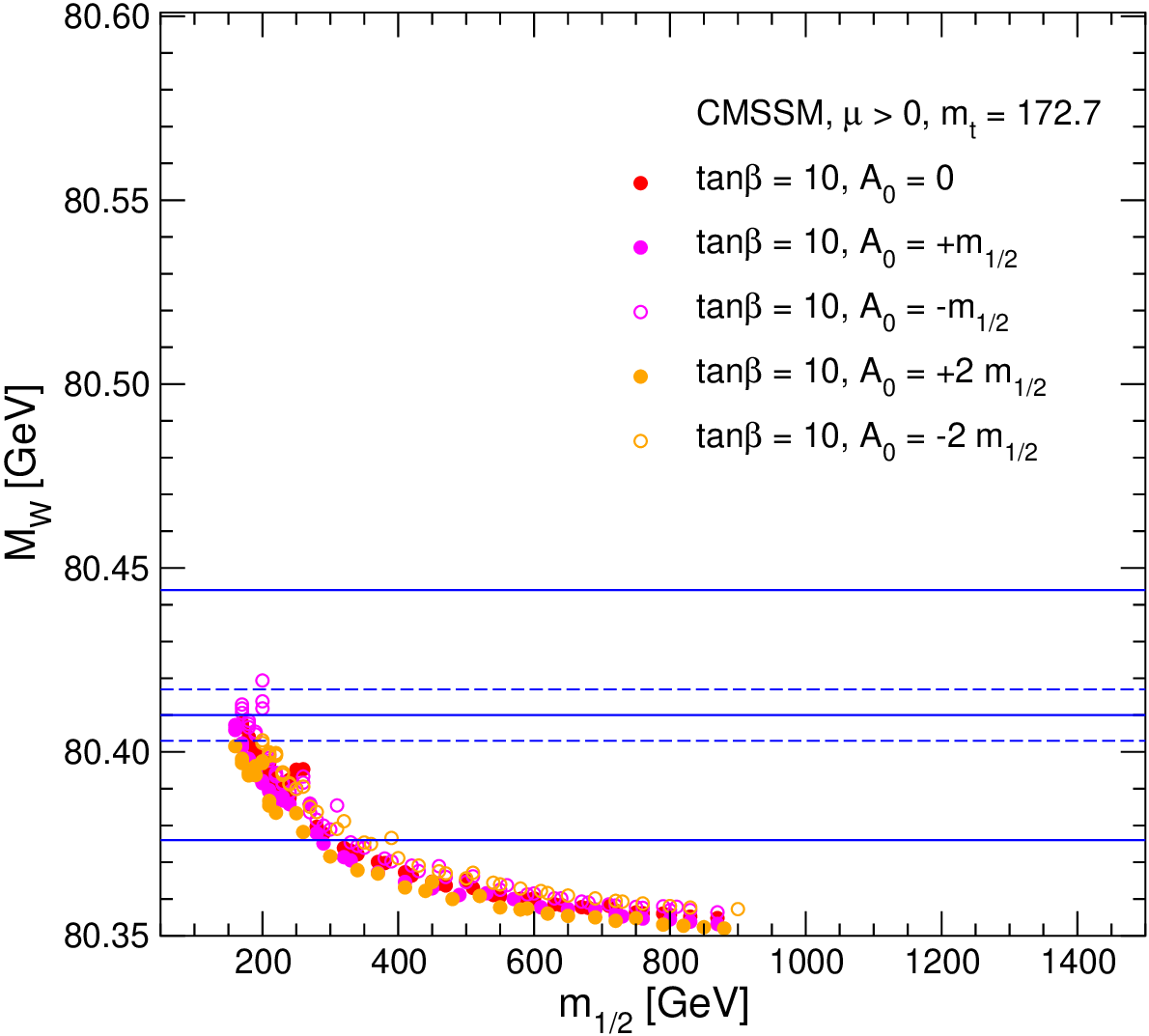}
\vspace{-0.5cm}
\caption{
\it The CMSSM predictions for $\MW$ as functions of $m_{1/2}$ for 
$\tb = 10$ for various $A_0$. The top quark mass has been set to 
$\mt = 178.0 \gev$ (left) and $\mt = 172.7 \gev$ (right). The
experimental measurements indicated in the plots are the previous one,
$\MW = 80.426 \pm 0.034 \gev$ (left) and the newer one,
$\MW = 80.410 \pm 0.032 \gev$ (right).
} 
\label{fig:MW}
\end{center}
\end{figure}

\begin{figure}[htb!]
\begin{center}
\vspace{-0.5cm}
\includegraphics[width=.45\textwidth,height=6cm]{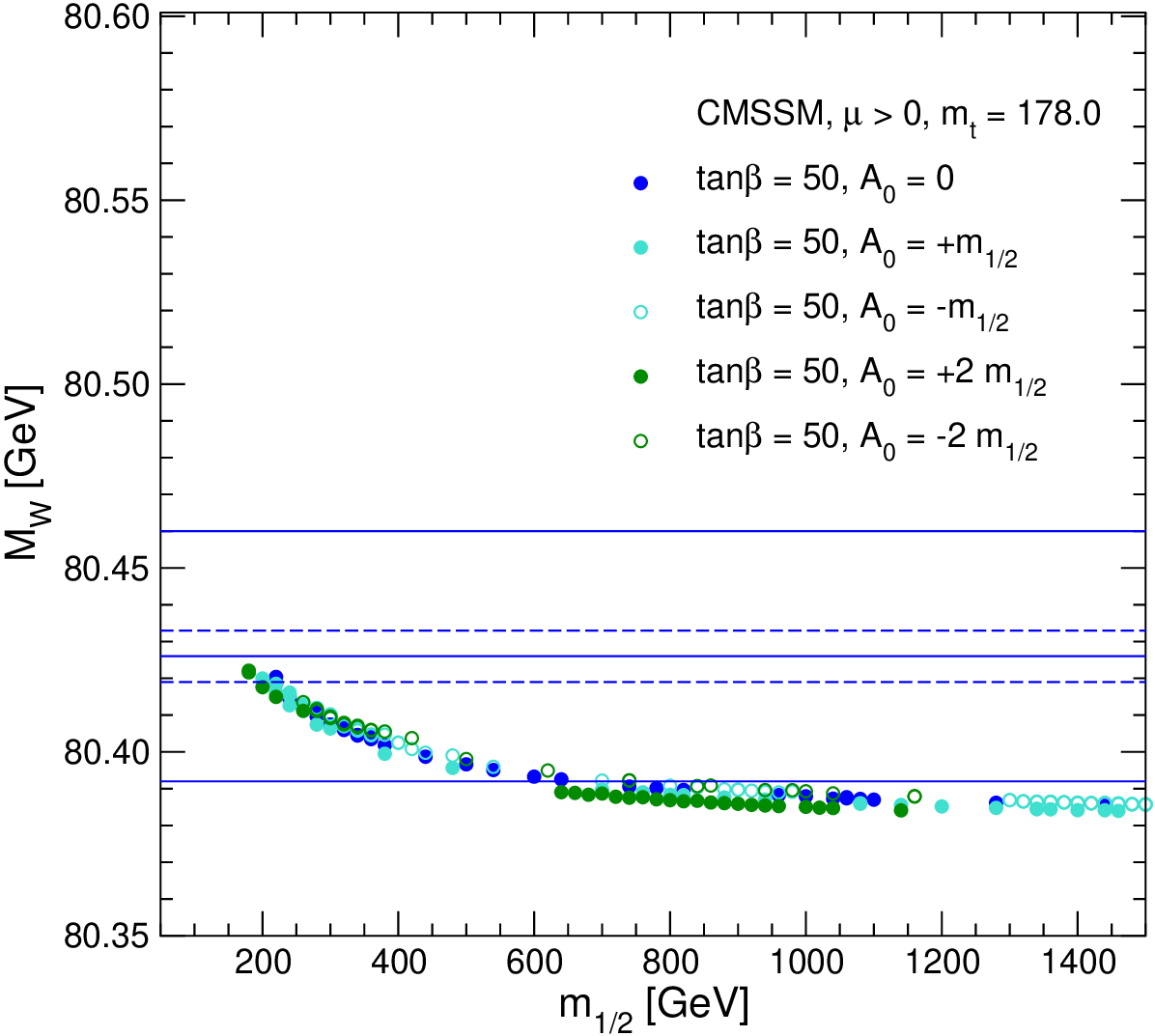}\hspace{1em}
\includegraphics[width=.45\textwidth,height=6cm]{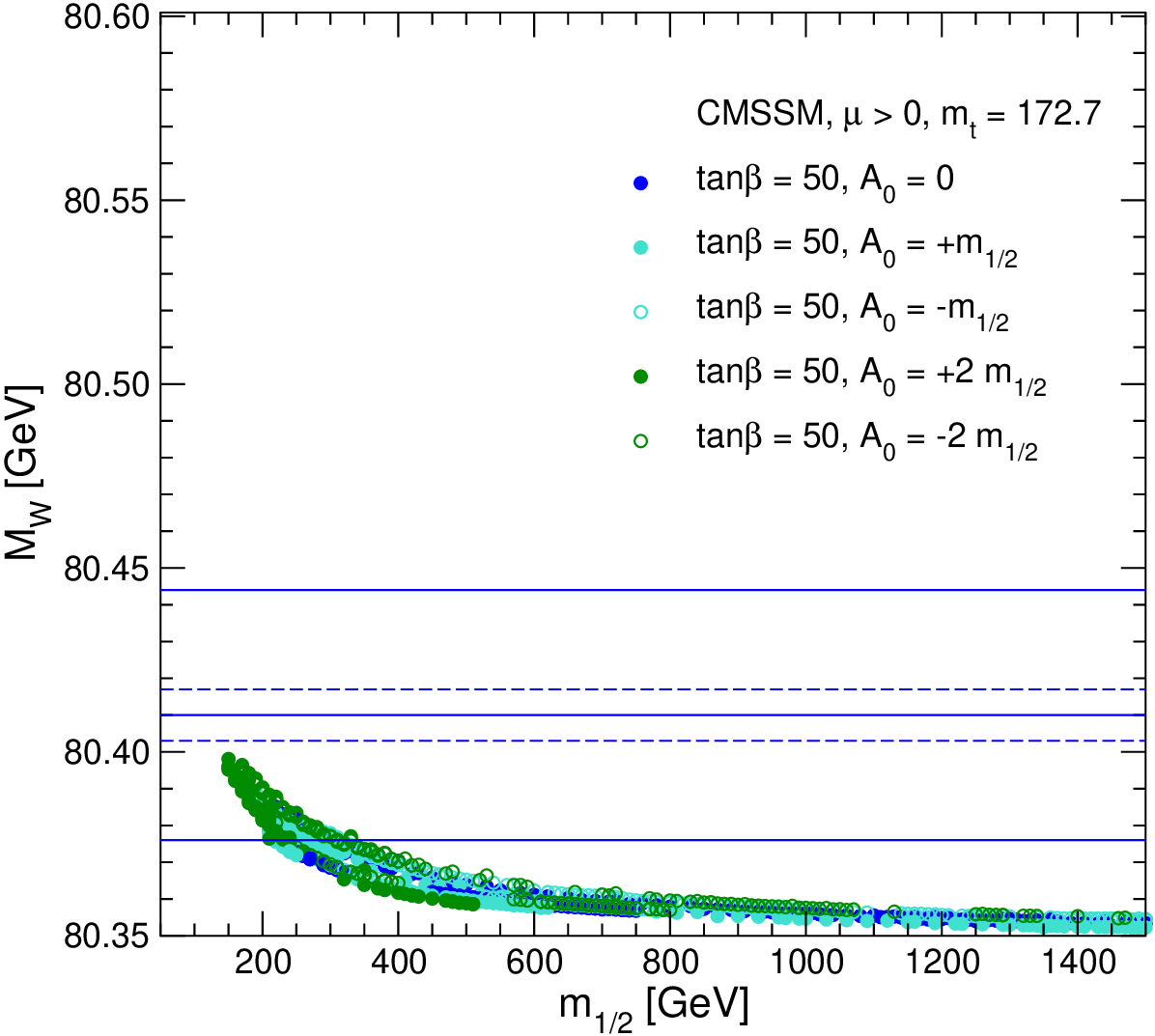}
\vspace{-0.5cm}
\caption{
\it Same as in \reffi{fig:MW}, but for $\tb = 50$.
} 
\label{fig:MWtb50}
\end{center}
\end{figure}

\begin{figure}[htb!]
\begin{center}
\vspace{-0.5cm}
\includegraphics[width=.45\textwidth,height=6cm]{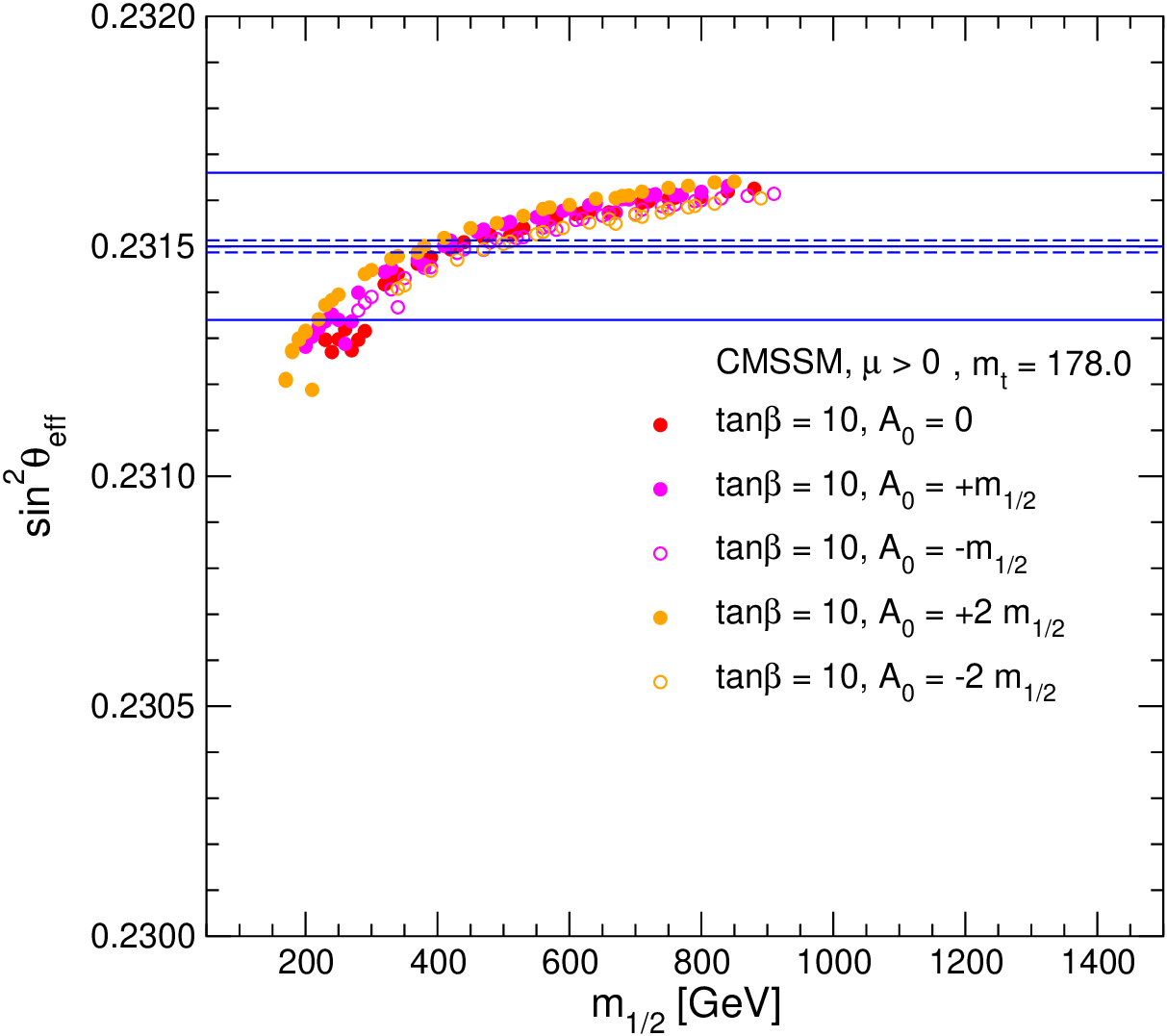}\hspace{1em}
\includegraphics[width=.45\textwidth,height=6cm]{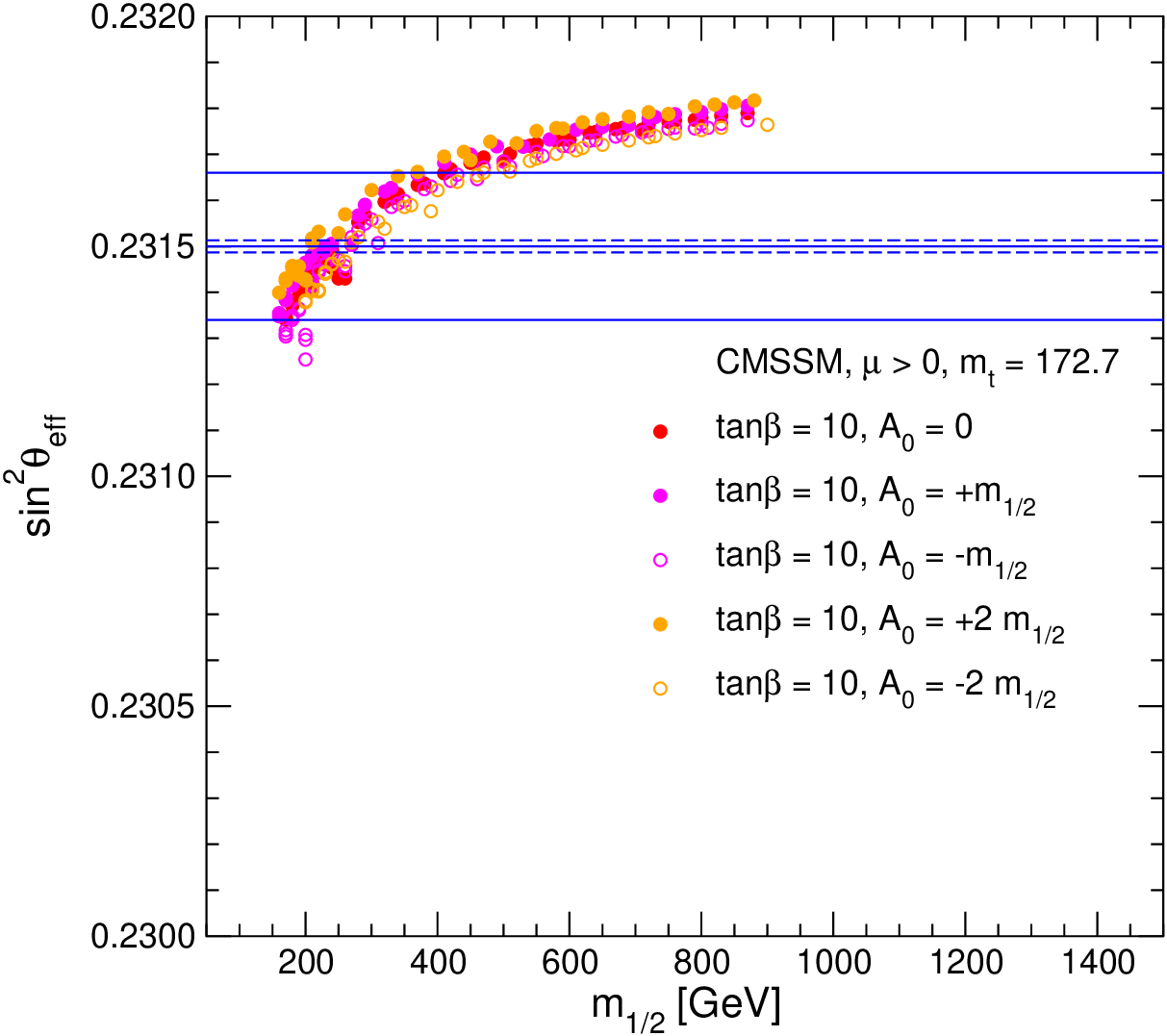}
\vspace{-0.5cm}
\caption{
\it The CMSSM predictions for $\sweff$ as functions of $m_{1/2}$ for 
$\tb = 10$ for various $A_0$. The top quark mass has been set to 
$\mt = 178.0 \gev$ (left) and $\mt = 172.7 \gev$ (right). 
} 
\label{fig:SW}
\end{center}
\end{figure}

\begin{figure}[htb!]
\begin{center}
%\vspace{-0.5cm}
\includegraphics[width=.45\textwidth,height=6cm]{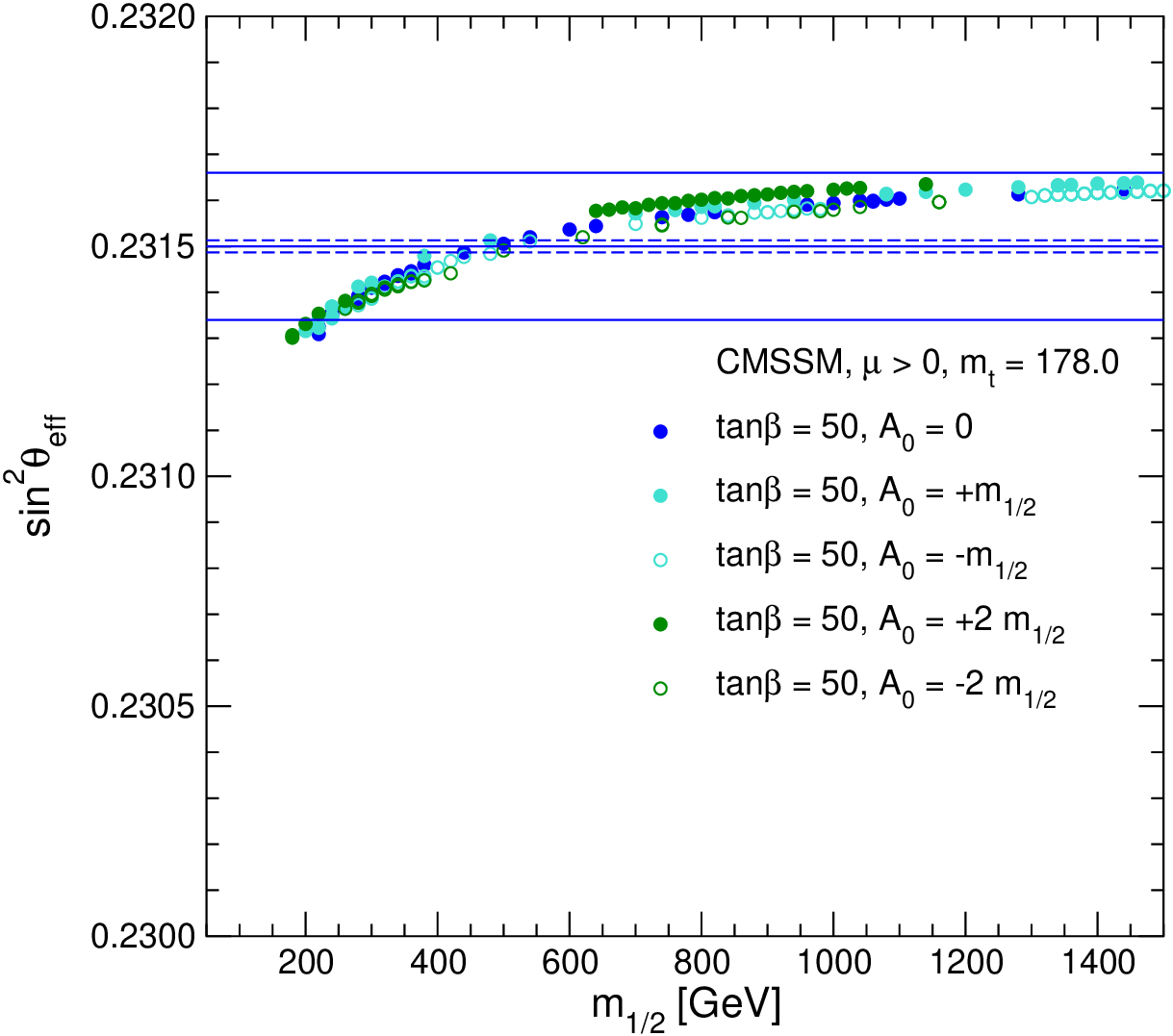}\hspace{1em}
\includegraphics[width=.45\textwidth,height=6cm]{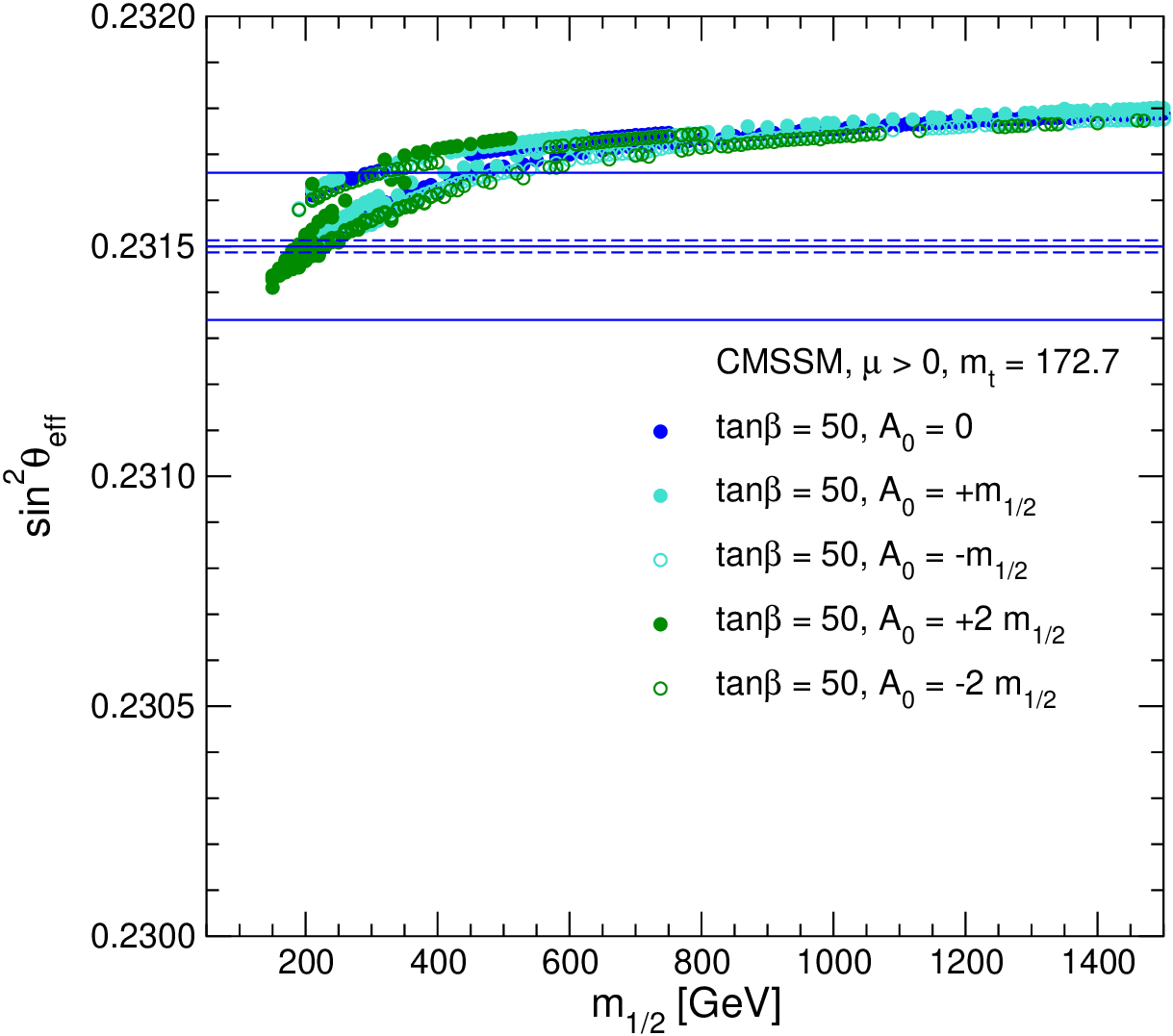}
\vspace{-0.5cm}
\caption{
\it Same as in \reffi{fig:SW}, but for $\tb = 50$.
} 
\label{fig:SWtb50}
\end{center}
\end{figure}

First, it drives the
SM prediction of $\MW$ and $\sweff$ further away from the
current experimental value~\footnote{Whereas $(g-2)_\mu$ and 
$\br(b \to s \ga)$ are 
little affected.}. This effect is shown in \reffis{fig:MW} -- \ref{fig:SWtb50}
for $\tb = 10, 50$.
In the right plots of \reffis{fig:MW} and \ref{fig:MWtb50} we have
also updated the 
experimental value of $\MW$. The change in the SM prediction elevates
the experimental discrepancy to about 1.5 $\sigma$, despite the change
in the preferred experimental range of $\MW$, which does not
compensate completely for the change in $\mt$. The net effect is
therefore to increase the favoured magnitude of the supersymmetric
contribution, i.e., to lower the preferred supersymmetric mass scale.
In the case of $\sweff$, the reduction in $\mt$ has increased the SM
prediction whereas the experimental value has not changed
significantly. Once again, the discrepancy with the SM has increased
to about 1.5 $\sigma$, and the preference for a small value of
$m_{1/2}$ has therefore also increased. 

%%%%%%%%%%%%%%%%%%%%%%%%% F I G U R E %%%%%%%%%%%%%%%%%%%%%%%%%%%%%%%%%%%%%%%%%
%%%%%%%%%%%%%%%%%%%%%%%%% F I G U R E %%%%%%%%%%%%%%%%%%%%%%%%%%%%%%%%%%%%%%%%%

%%%%%%%%%%%%%%%%%%%%%%%%% F I G U R E %%%%%%%%%%%%%%%%%%%%%%%%%%%%%%%%%%%%%%%%%
%%%%%%%%%%%%%%%%%%%%%%%%% F I G U R E %%%%%%%%%%%%%%%%%%%%%%%%%%%%%%%%%%%%%%%%%

%%%%%%%%%%%%%%%%%%%%%%%%% F I G U R E %%%%%%%%%%%%%%%%%%%%%%%%%%%%%%%%%%%%%%%%%
%%%%%%%%%%%%%%%%%%%%%%%%% F I G U R E %%%%%%%%%%%%%%%%%%%%%%%%%%%%%%%%%%%%%%%%%

%%%%%%%%%%%%%%%%%%%%%%%%% F I G U R E %%%%%%%%%%%%%%%%%%%%%%%%%%%%%%%%%%%%%%%%%
%%%%%%%%%%%%%%%%%%%%%%%%% F I G U R E %%%%%%%%%%%%%%%%%%%%%%%%%%%%%%%%%%%%%%%%%

\begin{figure}[htb!]
\begin{center}
%\vspace{-0.5cm}
\includegraphics[width=.45\textwidth,height=6cm]{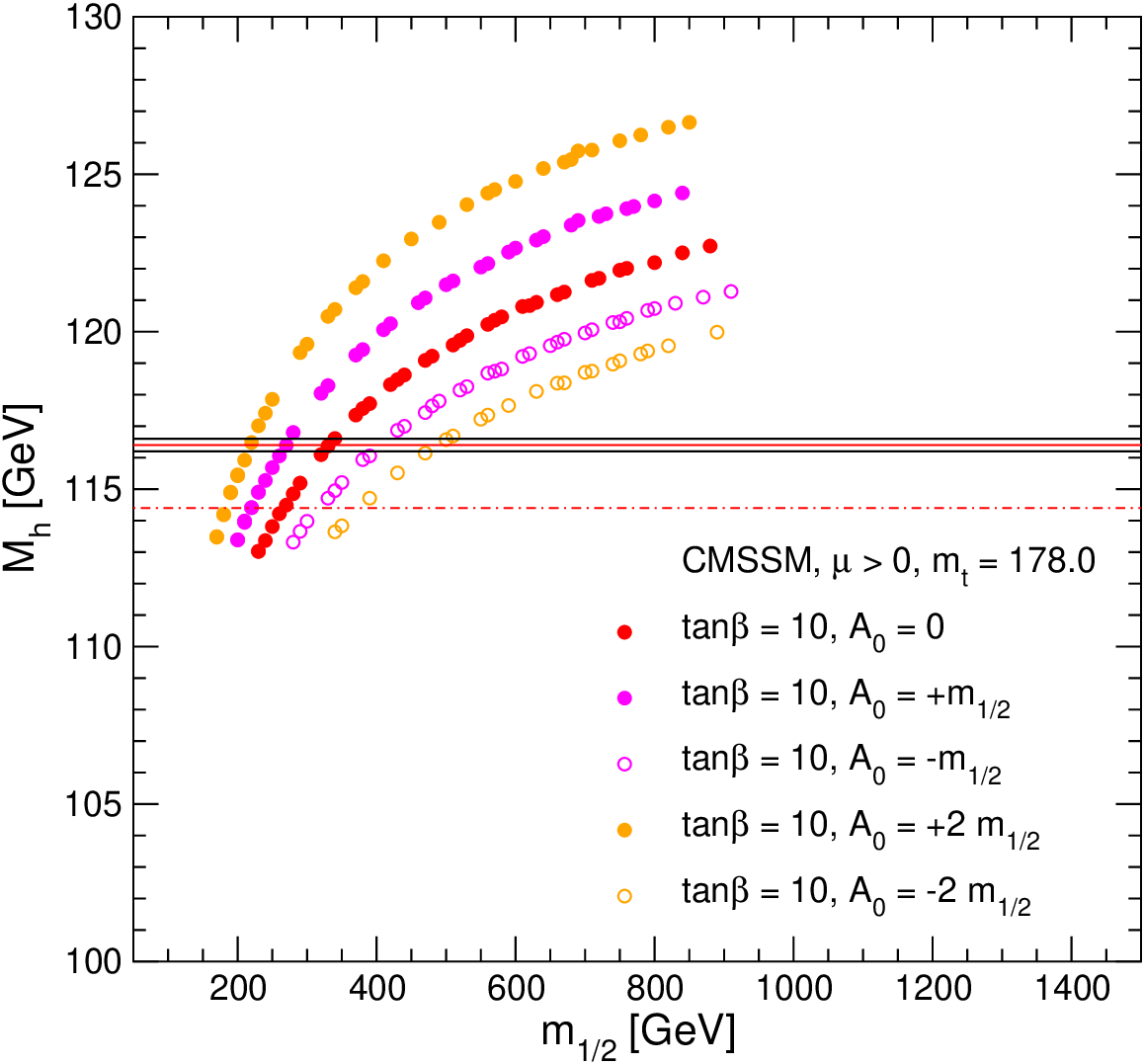}\hspace{1em}
\includegraphics[width=.45\textwidth,height=6cm]{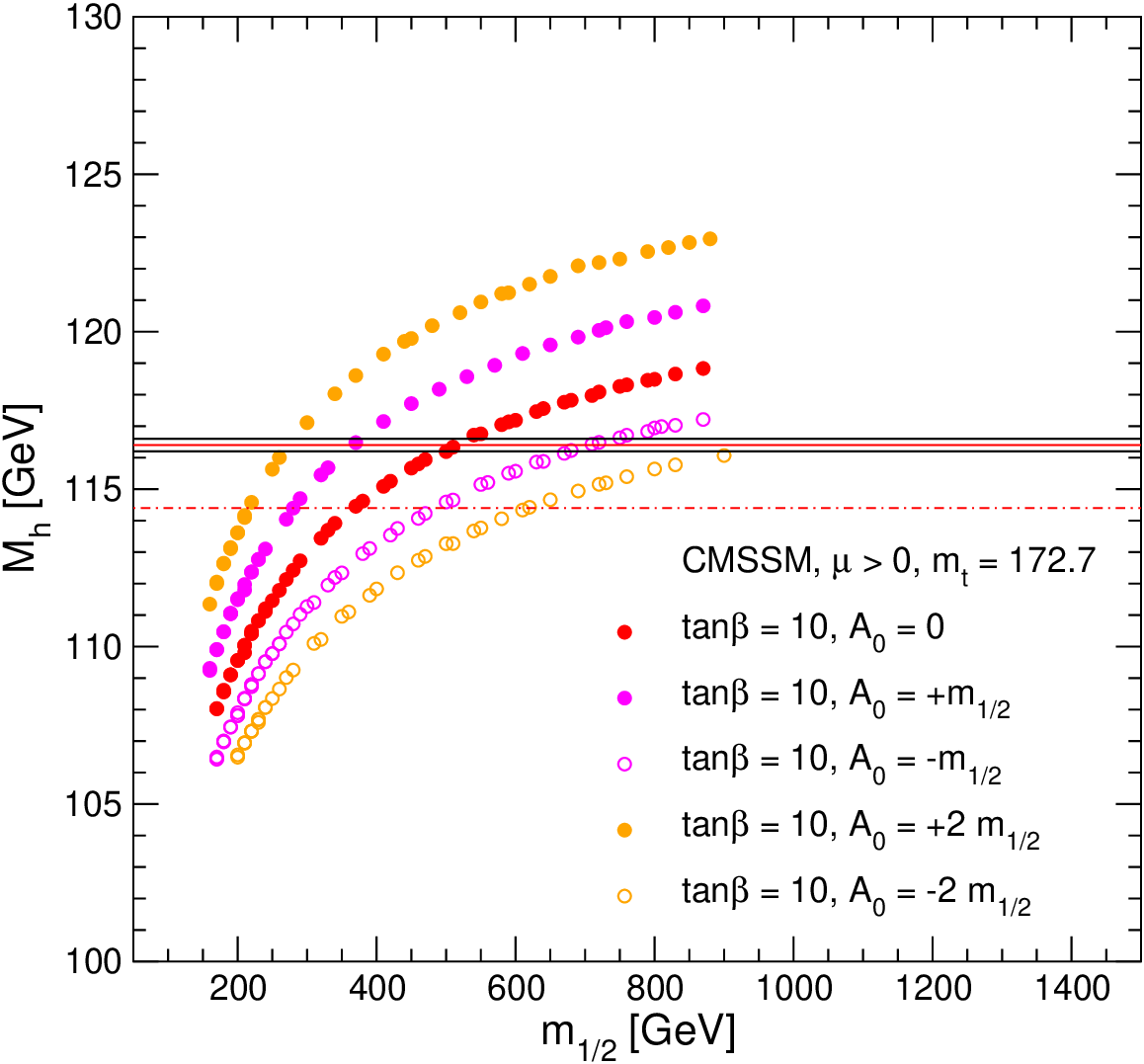}
\vspace{-0.5cm}
\caption{
\it The CMSSM predictions for $\Mh$ as functions of $m_{1/2}$ for 
$\tb = 10$ for various $A_0$. The top quark mass has been set to 
$\mt = 178.0 \gev$ (left) and $\mt = 172.7 \gev$ (right). 
A hypothetical LHC measurement is also shown, %in the right plot, 
namely $\Mh = 116.4 \pm 0.2 \gev$, 
as well as the present 95\%~C.L.\ exclusion limit of $114.4 \gev$.
} 
\label{fig:Mh}
\end{center}
\end{figure}

\begin{figure}[htb!]
\begin{center}
\vspace{-0.5cm}
\includegraphics[width=.45\textwidth,height=6cm]{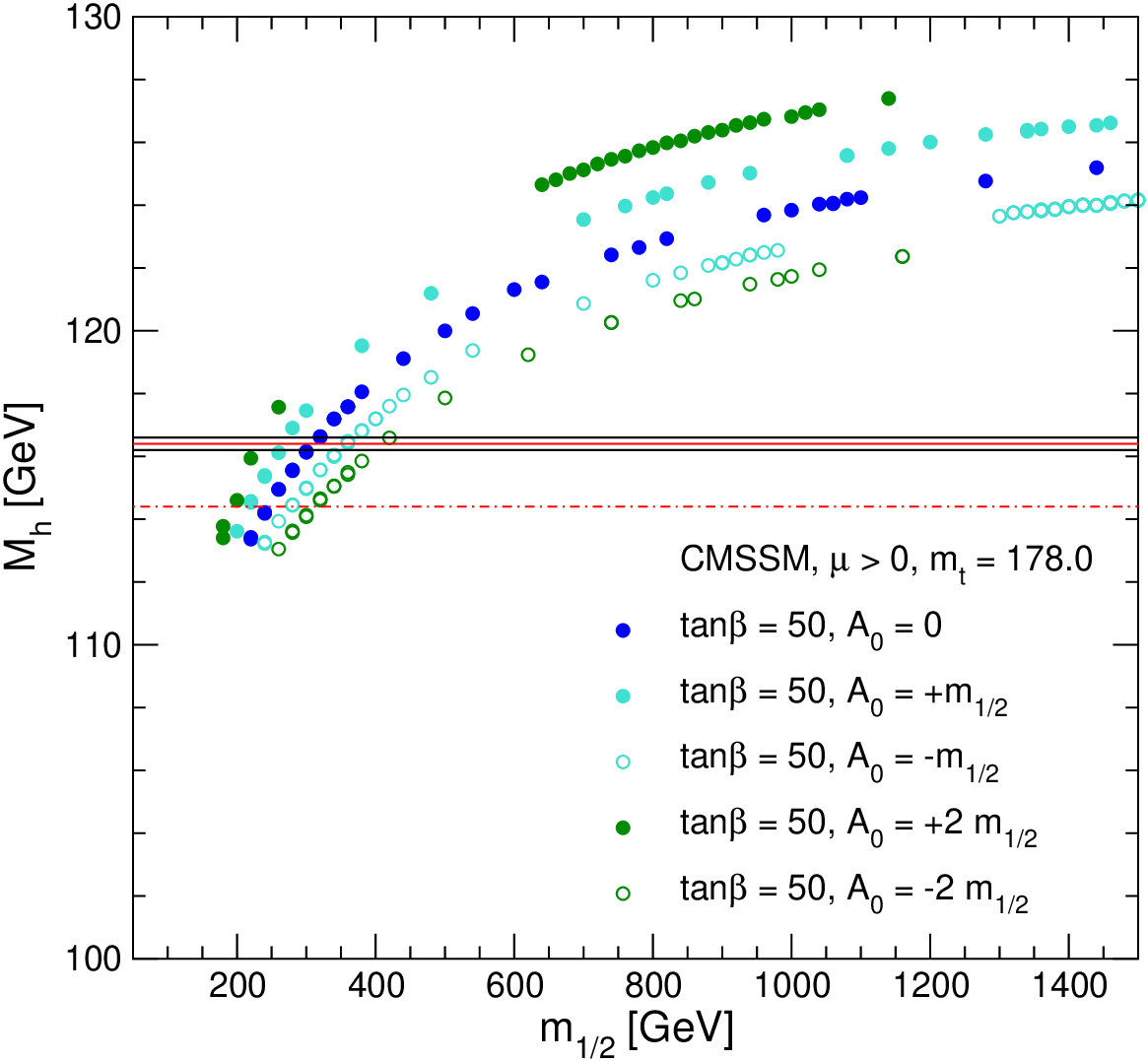}\hspace{1em}
\includegraphics[width=.45\textwidth,height=6cm]{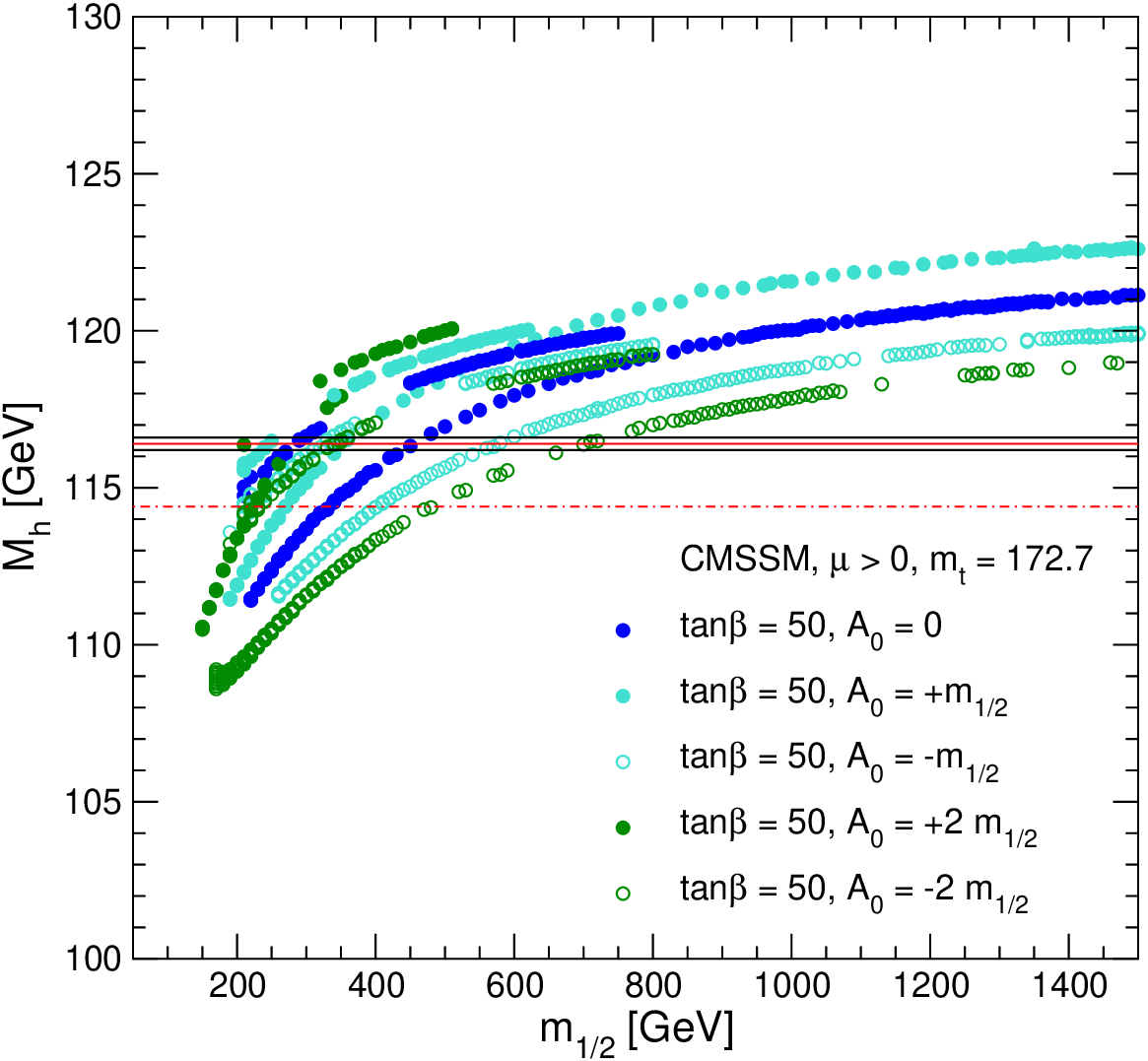}
\vspace{-0.5cm}
\caption{
\it Same as in \reffi{fig:Mh}, but for $\tb = 50$.
} 
\label{fig:Mhtb50}
\end{center}
\end{figure}

Secondly, the predicted value of the lightest Higgs boson mass in the
MSSM is lowered by the new $\mt$ value, see, e.g., \citere{tbexcl}.
The effects on the electroweak precision
observables of the downward shift in $\Mh$ are minimal, but
the LEP Higgs bounds~\cite{LEPHiggsSM,LEPHiggsMSSM} now impose a more
important constraint on the MSSM parameter space, notably on $m_{1/2}$.
This is visualized in \reffis{fig:Mh} and \reffi{fig:Mhtb50}, where we
show the results for 
$\tb = 10, 50$ for $\mt = 178.0 \gev$ (left plots) and $\mt = 172.7 \gev$
(right plots). 
A hypothetical LHC measurement is also shown, 
namely $\Mh = 116.4 \pm 0.2 \gev$, 
as well as the present 95\%~C.L.\ exclusion limit of $114.4 \gev$.
%
%%%%%%%%%%%%%%%%%%%%%%%%% F I G U R E %%%%%%%%%%%%%%%%%%%%%%%%%%%%%%%%%%%%%%%%%
%%%%%%%%%%%%%%%%%%%%%%%%% F I G U R E %%%%%%%%%%%%%%%%%%%%%%%%%%%%%%%%%%%%%%%%%
%
%%%%%%%%%%%%%%%%%%%%%%%%% F I G U R E %%%%%%%%%%%%%%%%%%%%%%%%%%%%%%%%%%%%%%%%%
%%%%%%%%%%%%%%%%%%%%%%%%% F I G U R E %%%%%%%%%%%%%%%%%%%%%%%%%%%%%%%%%%%%%%%%%
%
For $\tb = 10$ as shown in \reffi{fig:Mh}, with the lower $\mt$ value
for small $m_{1/2}$, a positive value of 
$A_0$ is needed in order to satisfy the LEP Higgs exclusion bounds.
For $\tb = 50$, see \reffi{fig:Mhtb50}, on the other hand, this effect
is much less severe.  
In our previous analysis, we rejected all parameter points for which {\tt 
FeynHiggs} yielded $\Mh < 113 \gev$. The best fit values in \citere{ehow3} 
corresponded to 
relatively small values of $\Mh$, a feature that is even more pronounced
for the new $\mt$ value. 
In view of all these effects, we have updated~\cite{ehow4} our old 
analysis of the phenomenological constraints on the supersymmetric mass 
scale $m_{1/2}$ in the CMSSM using the new, lower
value~% 
\footnote{
See also \citere{LCWS05ehow3}, where a lower bound of 
$\Mh > 111.4 \gev$ has been used.
}%
~of $\mt$
and including a $\chi^2$ contribution from $\Mh$.

%%%%%%%%%%%%%%%%%%%%%%%%%%%%%%%%%%%%%%%%%%%%%%%%%%%%%%%%%%%%%%%%%%%%%
%%%%%%%%%%%%%%%%%%%%%%%%%%%%%%%%%%%%%%%%%%%%%%%%%%%%%%%%%%%%%%%%%%%%%

\begin{figure}[htb!]
%\vspace{1em}
\begin{center}
%\vspace{-0.5cm}
\includegraphics[width=.45\textwidth,height=5.4cm]{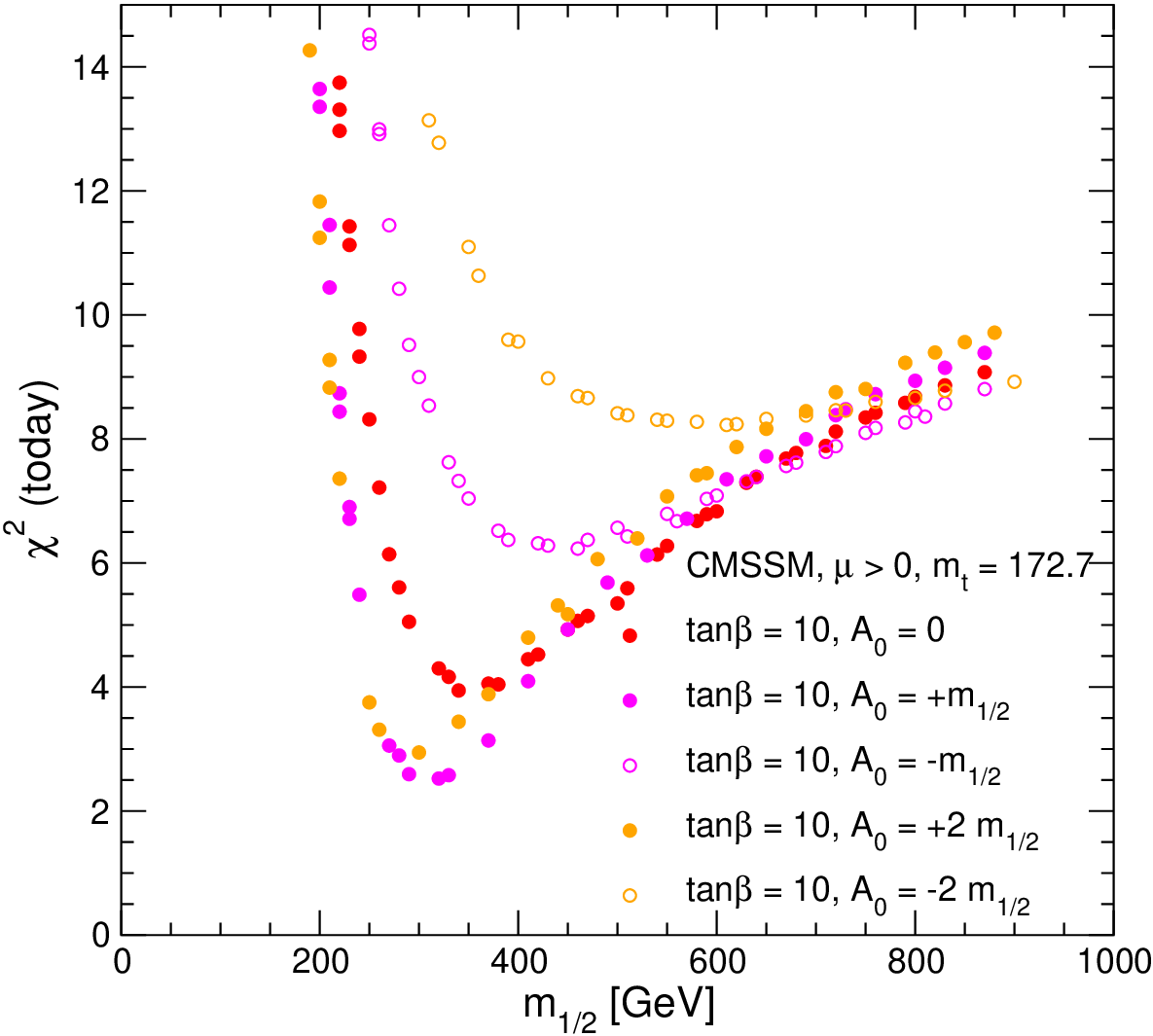}
\includegraphics[width=.45\textwidth,height=5.4cm]{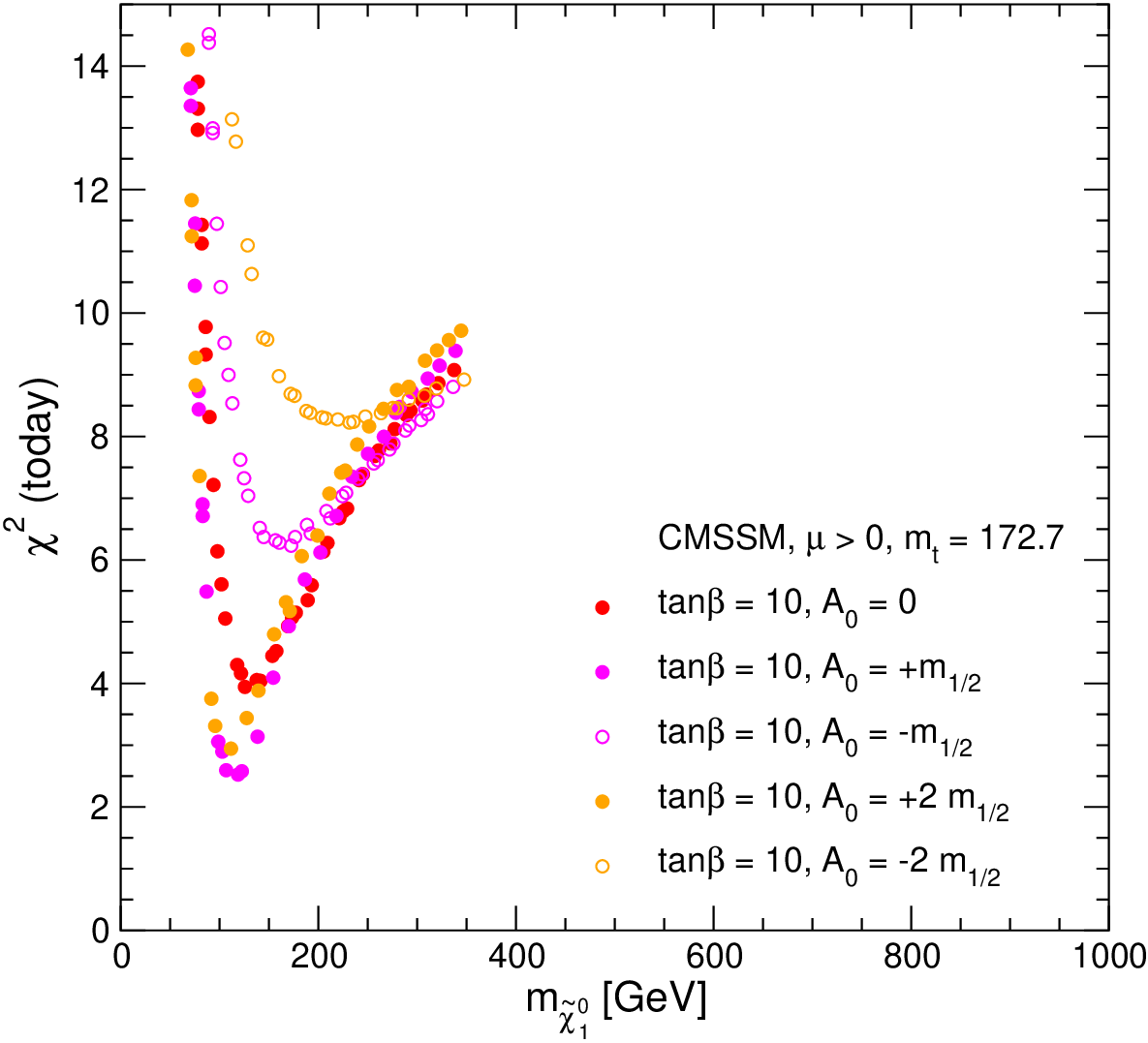}\\[1em]
\includegraphics[width=.45\textwidth,height=5.4cm]{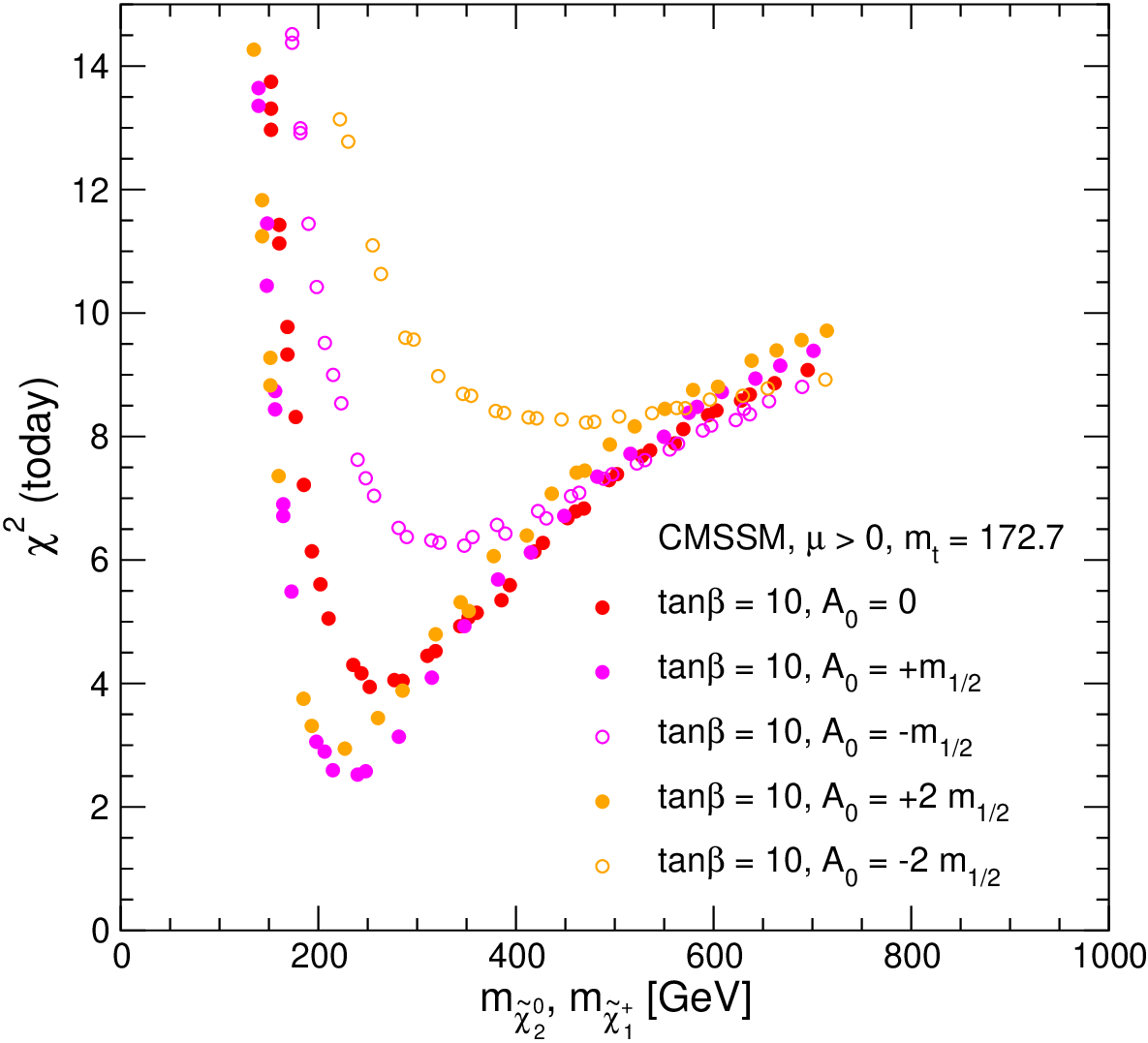}
\includegraphics[width=.45\textwidth,height=5.4cm]{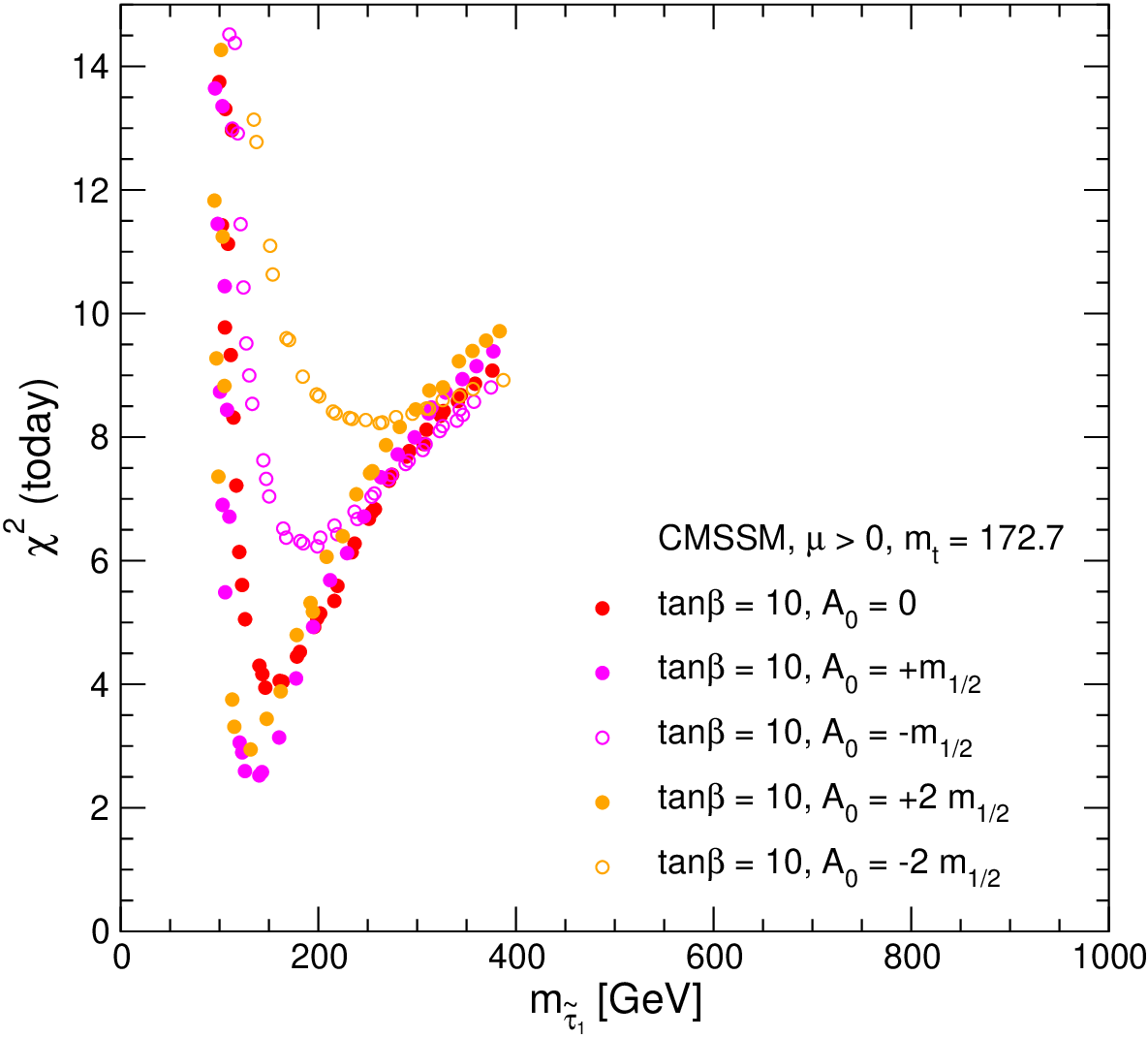}\\[1em]
\includegraphics[width=.45\textwidth,height=5.4cm]{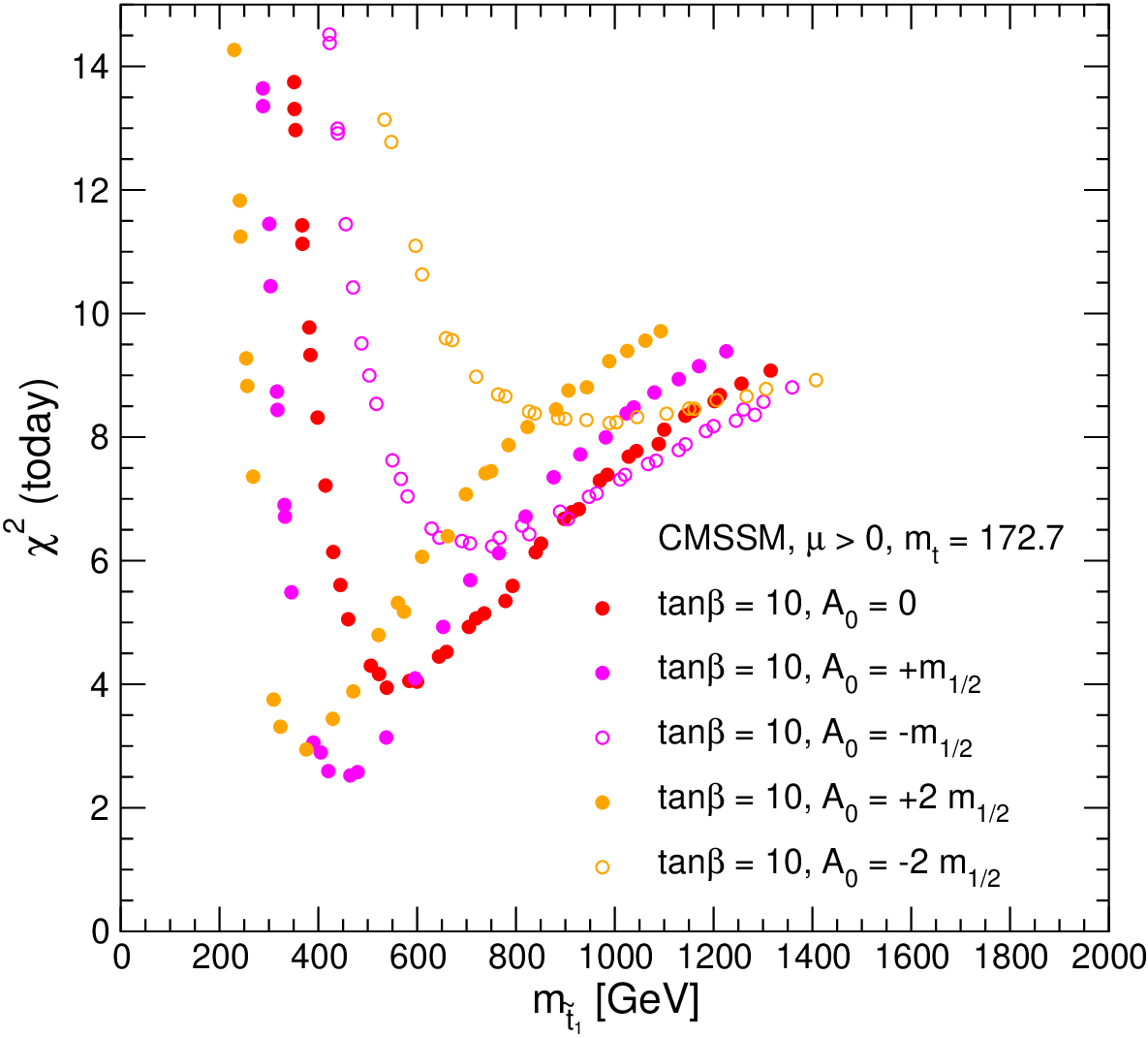}
\includegraphics[width=.45\textwidth,height=5.4cm]{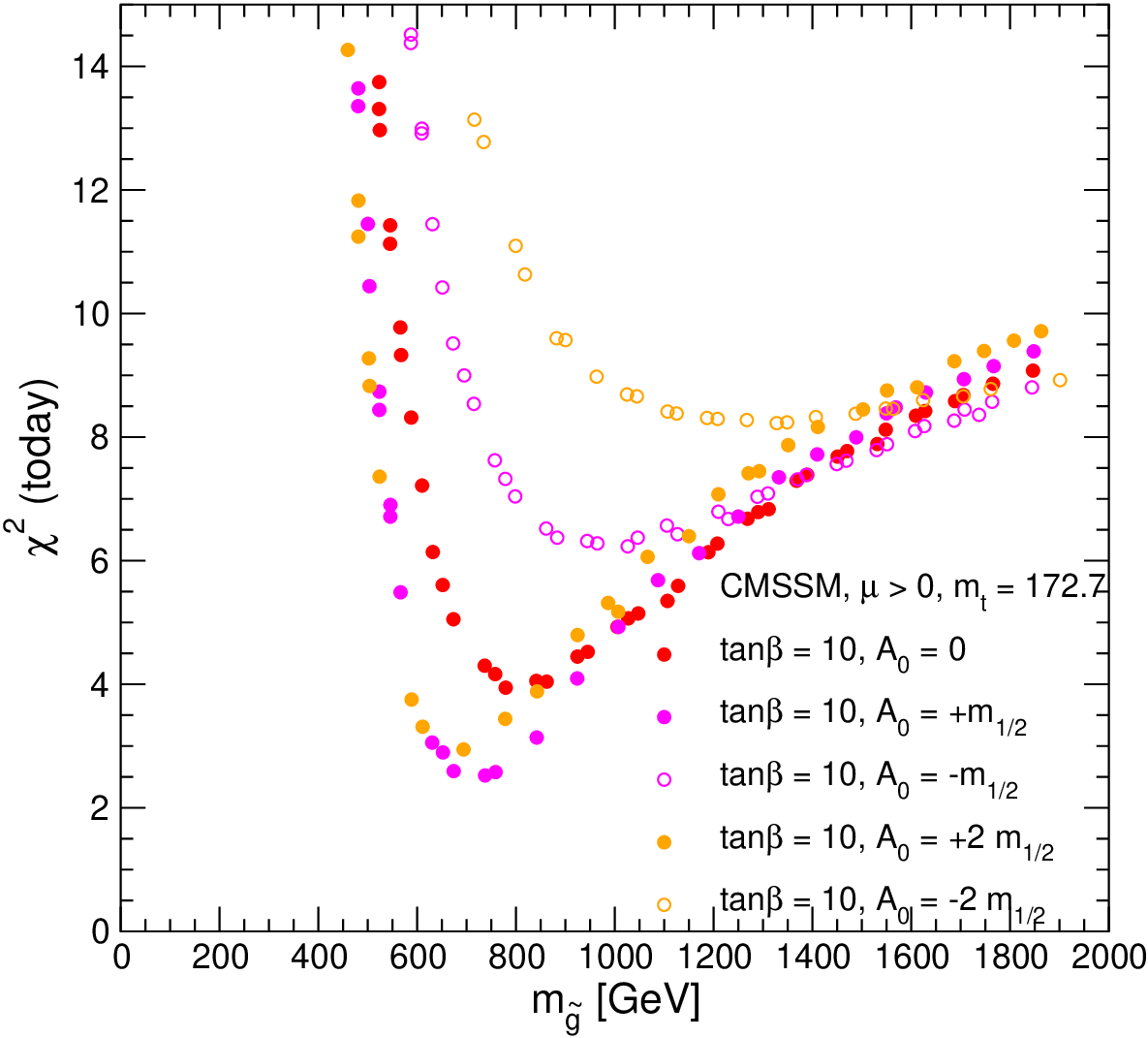}
\caption{
\it The combined likelihood function $\chi^2$ for the electroweak
observables $\MW$, $\sweff$,
$(g - 2)_\mu$, ${\rm BR}(b \to s \ga)$, and $\Mh$
evaluated in the CMSSM for $\tb = 10$,
$\mt = 172.7 \pm 2.9 \gev$ and various discrete values of $A_0$, with
$m_0$ then chosen to yield the central value of the relic neutralino 
density
indicated by WMAP and other observations. We display the $\chi^2$
function for (a) $m_{1/2}$, (b) $\mneu{1}$, 
(c) $\mneu{2}, \mcha{1}$, (d) $\mstaue$, (e) $\mste$ and 
(f) $\mgl$~\cite{ehow4}.
} 
\label{fig:newmt10}
\end{center}
\vspace{-2em}
\end{figure}

We now present the updated results~\cite{ehow4} for the $\chi^2$~fit, 
which includes the $\chi^2$ contribution for $\Mh$ for 
$\mt = 172.7 \pm 2.9 \gev$.%
\footnote{
The results for $\mt = 172.5 \pm 2.3 \gev$ are very similar, see also
\refse{sec:mtfuture}
}%
~As seen in the first panel of \reffi{fig:newmt10}, the qualitative
feature observed in \citere{ehow3} of a pronounced minimum in $\chi^2$ at
$m_{1/2}$ for $\tb = 10$ is also present for the new value of
$\mt$. However, the $\chi^2$ curve now depends more strongly on the value 
of $A_0$, corresponding to its strong impact on $\Mh$.
Values of $A_0/m_{1/2} < -1$ are disfavoured at the 
$\sim$ 90\%~C.L., essentially because of their lower $\Mh$ values (see
\reffi{fig:Mh}), but 
$A_0/m_{1/2} = 2$ and~1 give equally good fits and descriptions of the
data. The old best fit point in \citere{ehow3} had $A_0/m_{1/2} = -1$, but
there all $A_0/m_{1/2}$ gave a similarly good description of the experimental
data. The minimum $\chi^2$ 
value is about~2.5. This is somewhat higher than the result in
\citere{ehow3}, but still represents a good overall fit to the 
experimental data. 
The rise in the minimum value of $\chi^2$, compared to \citere{ehow3},
is essentially a consequence of the lower experimental central value of
$\mt$, and the consequent greater
impact of the LEP constraint on $\Mh$~\cite{LEPHiggsSM,LEPHiggsMSSM}.
In the cases of the observables $\MW$ and $\sweff$, a smaller value of 
$\mt$ induces a preference for a smaller value of $m_{1/2}$, but the
opposite is true for the Higgs mass bound. The rise in the
minimum value of $\chi^2$ reflects the correspondingly increased 
tension between the electroweak precision observables and the $\Mh$
constraint. 

\begin{figure}[htb!]
%\vspace{1em}
\begin{center}
%\vspace{-0.5cm}   
\includegraphics[width=.45\textwidth,height=5.5cm]{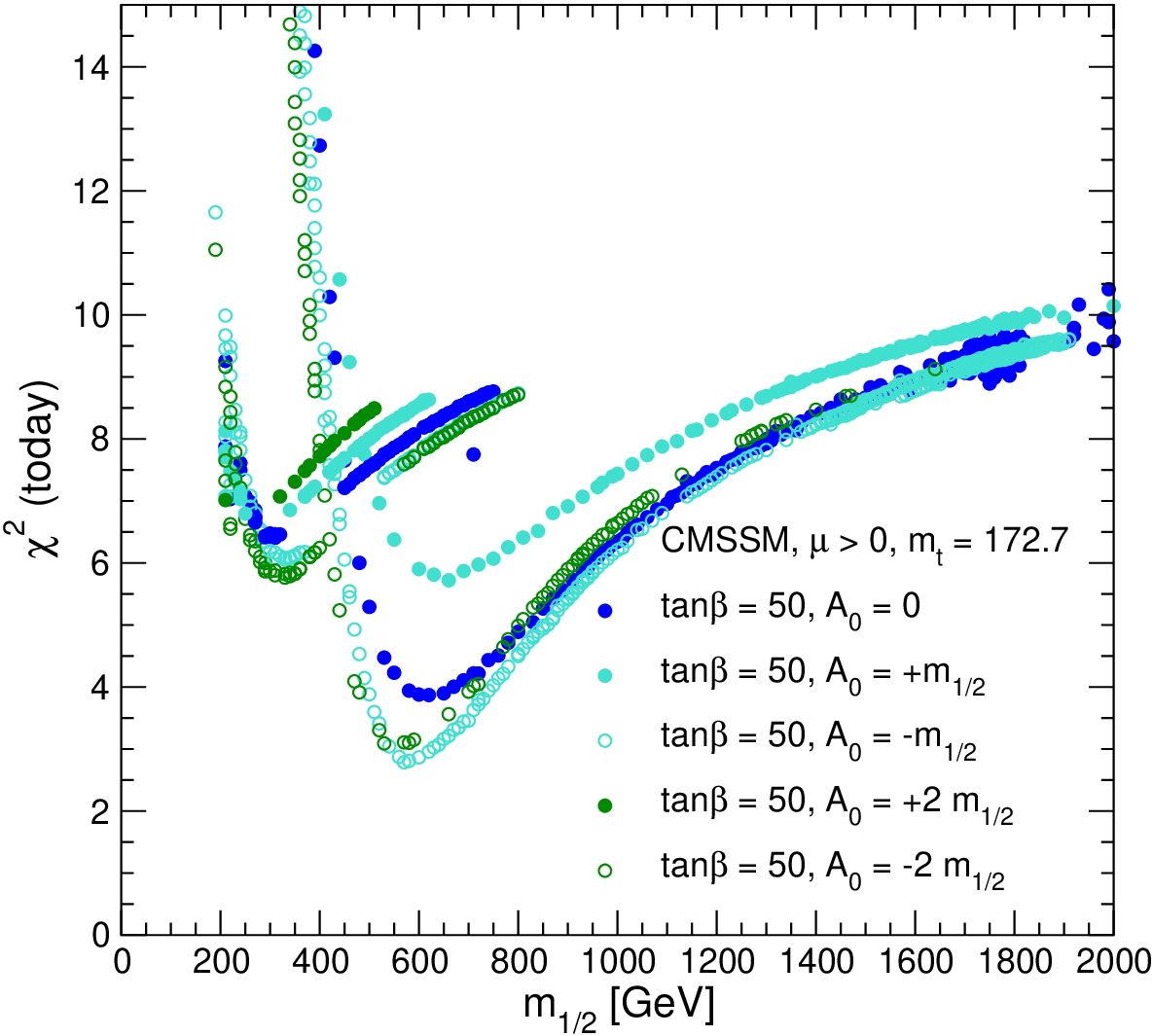}
\includegraphics[width=.45\textwidth,height=5.5cm]{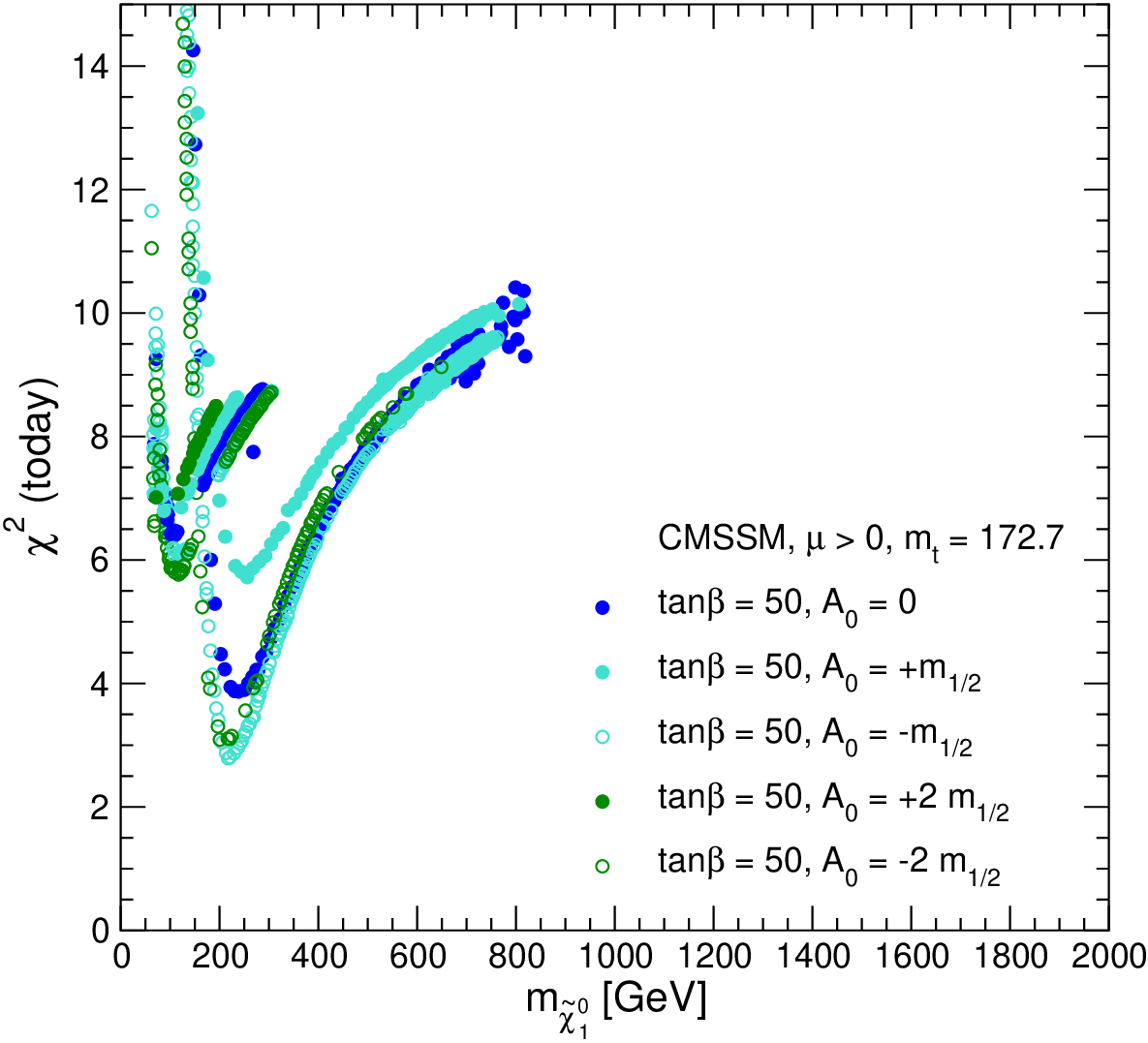}\\[1em]
\includegraphics[width=.45\textwidth,height=5.5cm]{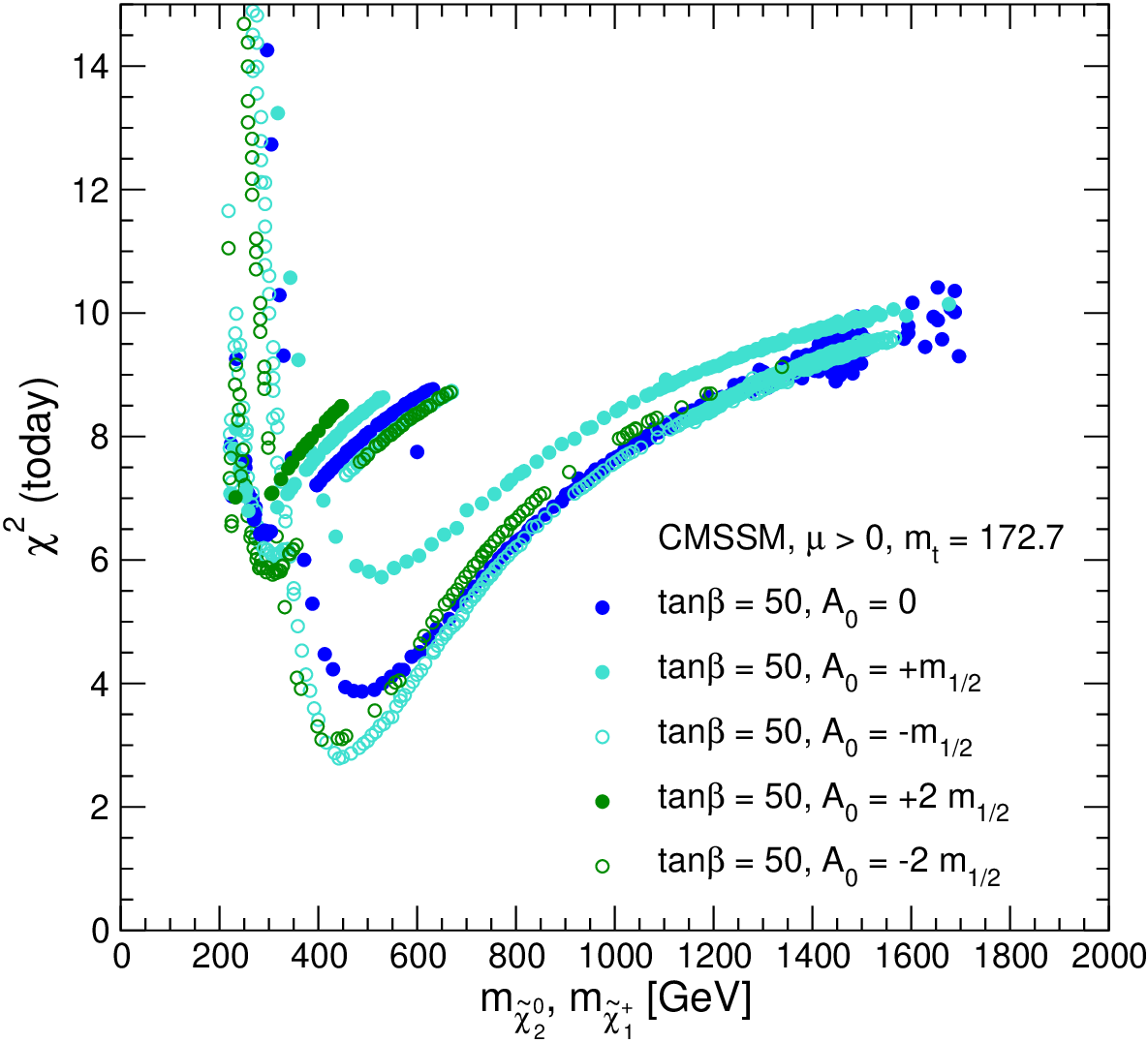}
\includegraphics[width=.45\textwidth,height=5.5cm]{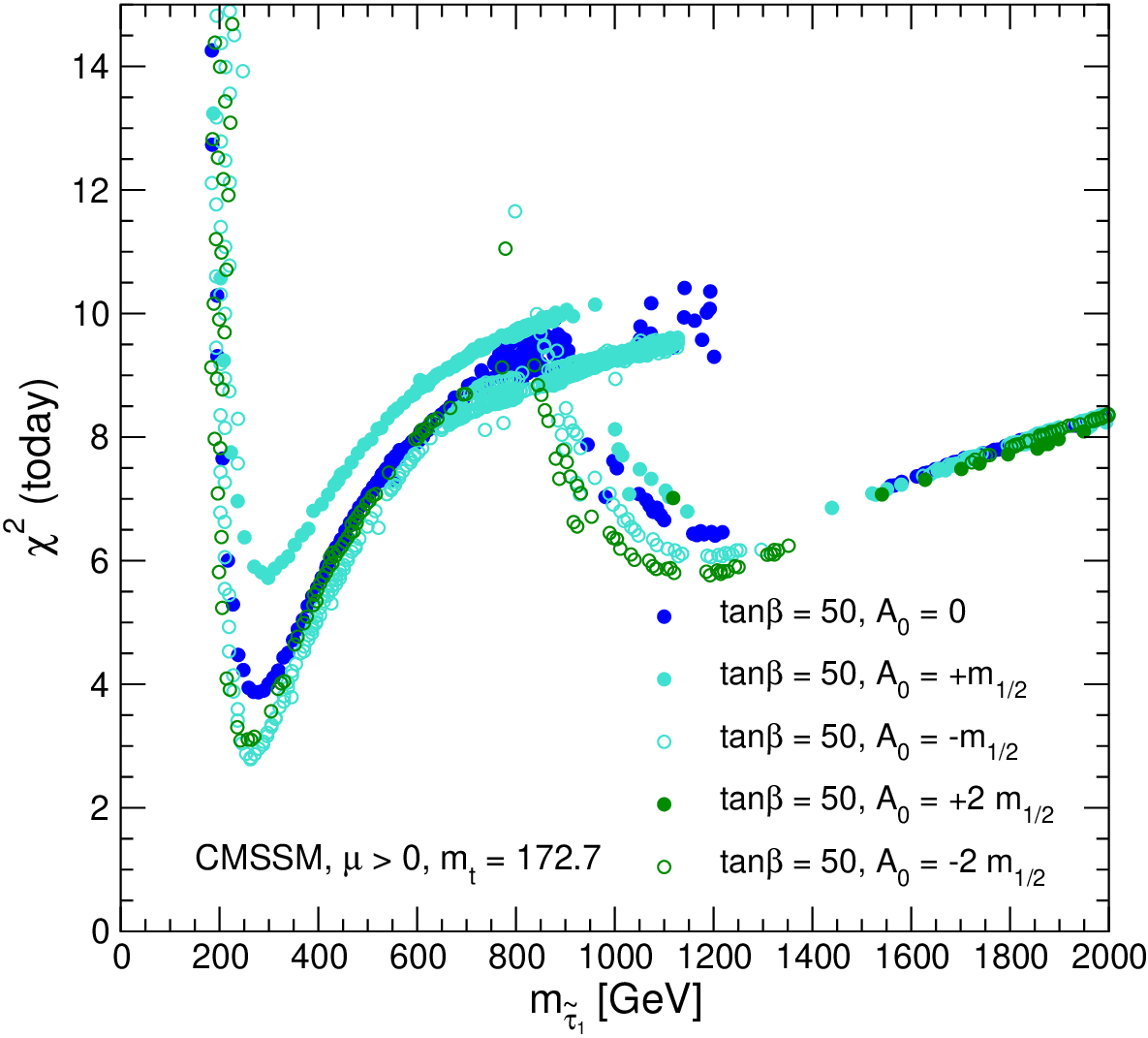}\\[1em]
\includegraphics[width=.45\textwidth,height=5.5cm]{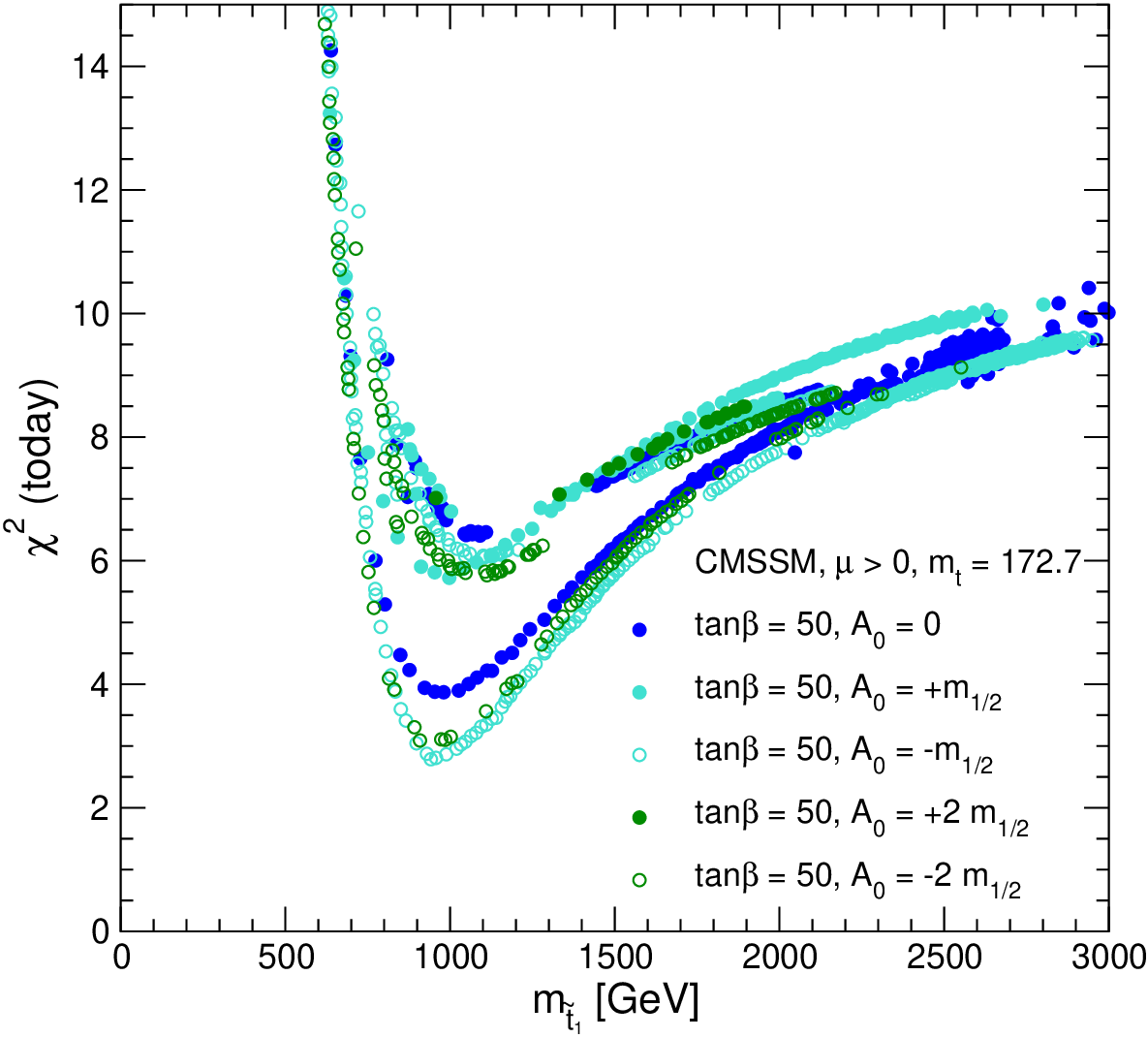}
\includegraphics[width=.45\textwidth,height=5.5cm]{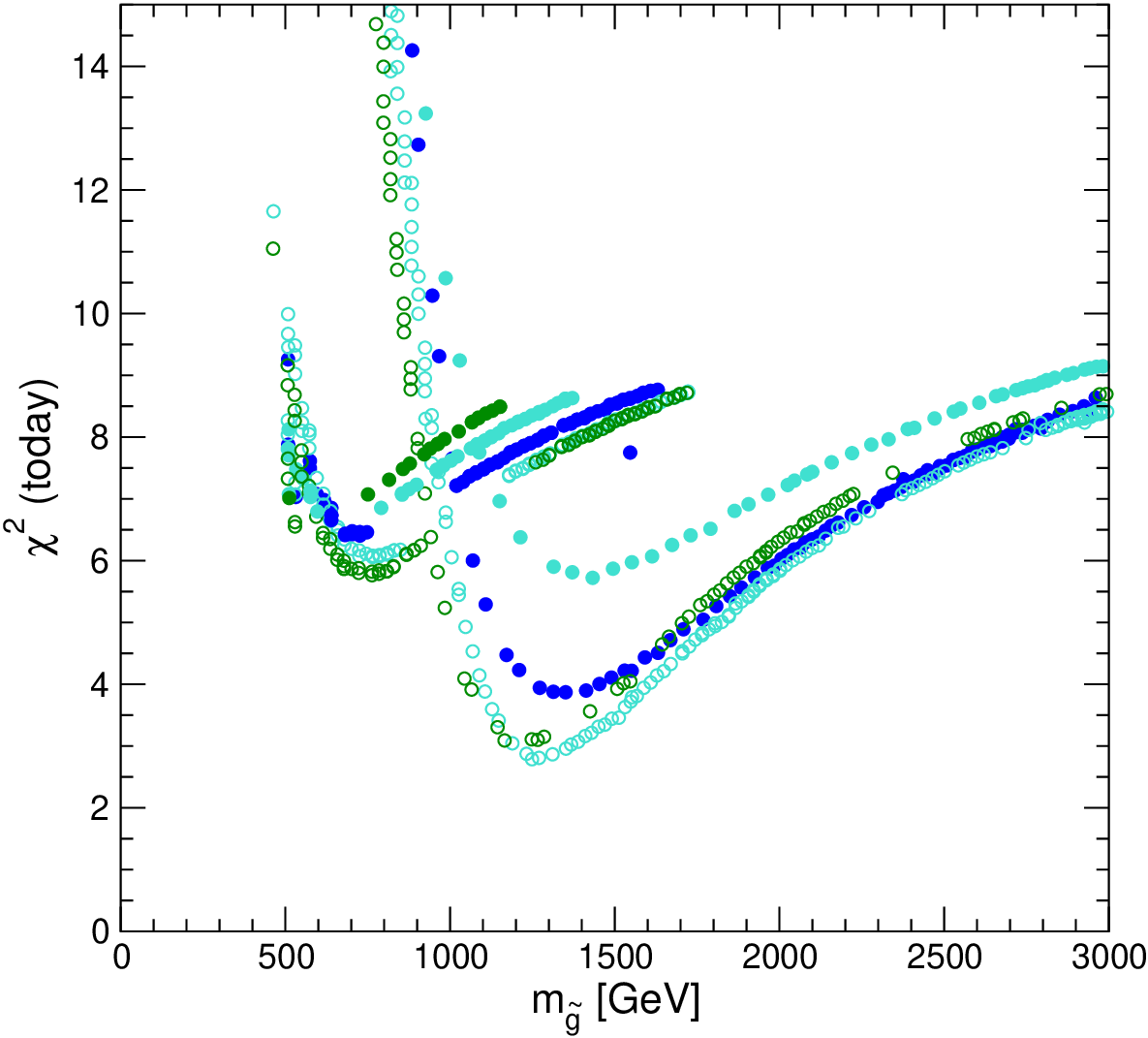}
\caption{
\it
As in Fig.~\protect\ref{fig:newmt10}, but now for $\tb = 50$.
}
\label{fig:newmt50}
\end{center}
%\vspace{-3em}
\end{figure}

A breakdown of the contributions to $\chi^2$ from the 
different observables can be found for some example points
in Table~\ref{tab:chi2}. We concentrate here on parameter sets
with relatively bad fit qualities that either have large $m_{1/2}$
values or lie in the focus-point region (see below).
One can see that, for large $m_{1/2}$ values, $(g-2)_\mu$ always gives 
the dominant contribution. However, with the new lower experimental value of
$\mt$ also $\MW$ and $\sweff$ give a substantial contribution, adding up 
to more than 50\% of the $(g-2)_\mu$ contribution. On the other hand, 
$\Mh$ and $\br (b \to s \ga)$ make negligible contributions to
$\chi^2$ at these points. 
As seen from the example shown in the last line of the Table, focus points
may yield similar results for the electroweak precision observables as
in the SM, resulting in a relatively high 
$\chi^2$~value. This region is mostly disfavoured at the 
$\sim 90\%$~C.L.\ level, as also seen in \reffi{fig:newmt50}.

%%%%%%%%%%%%%%%%%%%%%% T A B L E %%%%%%%%%%%%%%%%%%%%%%%%%%%%%%%%%%%%%%%%%
\begin{table}[tbh!]
\renewcommand{\arraystretch}{1.5}
\BC
\begin{tabular}{|c|c|c|c||c||c|c|c|c|c|}
\hline\hline
$\tb$ & $m_{1/2}$ & $m_0$ & $A_0$ & $\chi^2_{\rm tot}$ & 
$\MW$ & $\sweff$ & $(g-2)_\mu$ & $\br(b \to s \ga)$ & $\Mh$ \\ \hline\hline
10 & 880 & 270 & 1760 & 9.71 & 2.29 & 1.28 & 6.14 & 0.01 & 0 \\ \hline
50 & 1910 & 1500 & -1910 & 9.61 & 2.21 & 1.11 & 6.29 & 0.01 & 0 \\ \hline
50 & 800 & 2970 & -800 & 8.73 & 1.92 & 0.72 & 6.05 & 0.04 & 0 \\
\hline\hline
\end{tabular}
\EC
\renewcommand{\arraystretch}{1}
\caption{\it Breakdown of $\chi^2$ contributions from the different 
precision observables to $\chi^2_{\rm tot}$ for some example points. 
All masses are in GeV. The last row
is representative of the focus-point region.
}
\label{tab:chi2}
\end{table}
%%%%%%%%%%%%%%%%%%%%%% T A B L E %%%%%%%%%%%%%%%%%%%%%%%%%%%%%%%%%%%%%%%%%

%%%%%%%%%%%%%%%%%%%%%%%%% F I G U R E %%%%%%%%%%%%%%%%%%%%%%%%%%%%%%%%%%%%%%%%%
%%%%%%%%%%%%%%%%%%%%%%%%% F I G U R E %%%%%%%%%%%%%%%%%%%%%%%%%%%%%%%%%%%%%%%%%

The remaining panels of \reffi{fig:newmt10} update our old
analyses~\cite{ehow3} of the $\chi^2$ functions for various sparticle 
masses within
the CMSSM, namely the lightest neutralino $\neu{1}$, the second-lightest
neutralino $\neu{2}$ and the (almost degenerate) lighter chargino
$\cha{1}$, the lightest slepton which is the lighter stau 
$\Staue$, the lighter stop squark $\Stope$, and the gluino
$\gl$. Reflecting the behaviour of the global $\chi^2$ function
in the first panel of \reffi{fig:newmt10}, the changes in the optimal
values of the sparticle masses are not large.
The 90\% C.L.\ upper bounds on the particle masses are nearly
unchanged compared to the results for $\mt = 178.0 \pm 4.3 \gev$ given in
\citere{ehow3}.

%%%%%%%%%%%%%%%%%%%%%%%%% F I G U R E %%%%%%%%%%%%%%%%%%%%%%%%%%%%%%%%%%%%%%%%%
%%%%%%%%%%%%%%%%%%%%%%%%% F I G U R E %%%%%%%%%%%%%%%%%%%%%%%%%%%%%%%%%%%%%%%%%

The corresponding results for the case $\tb = 50$ are shown in
\reffi{fig:newmt50}.  We see in panel (a) that the minimum value of
$\chi^2$ for the fit with $m_t = 172.7 \pm 2.9 \gev$ is larger by about a unit
than in our previous analysis with $m_t = 178.0 \pm 4.3 \gev$.  Because of
the rise in $\chi^2$ for the $\tb = 10$ case, however, the minimum values
of $\chi^2$ are now very similar for the two values of $\tb$ shown here. The
dip in the $\chi^2$ function for $\tb = 50$ is somewhat steeper than in
the previous analysis, since the high values of $m_{1/2}$ are slightly
more disfavoured due to their $\MW$ and $\sweff$ values. The best fit
values of $m_{1/2}$ are very similar to their previous values. The
preferred values of the sparticle masses are shown in the remaining panels
of \reffi{fig:newmt50}.

We note one novel feature, namely the appearance of a group of points with
moderately high $\chi^2$ that have relatively small 
$m_{1/2} \sim 200-800 \gev$.
These points have relatively large values of $m_0$, as reflected in the
relatively large values of $\mstaue$ and $\mste$
seen in panels (d) and (e) of \reffi{fig:newmt50}. These points are
located in the focus-point region of the $(m_{1/2}, m_0)$
plane~\cite{focus}, where the LSP has a larger Higgsino content, whose
enhanced annihilation rate brings the relic
density down into the range allowed by WMAP. These points have a 
$\De\chi^2$ of at least 3.5, so most of them are excluded at the
90\%~C.L. 

Taken at face value, the preferred ranges for the sparticle masses shown
in \reffis{fig:newmt10} and \ref{fig:newmt50} are quite
encouraging for both the LHC and the ILC. The gluino and squarks lie
comfortably within the early LHC discovery range, and several
electroweakly-interacting sparticles would be accessible to ILC(500)
(the ILC running at $\sqrt{s} = 500 \gev$). This is the case, in
particular, for the 
${\tilde \chi}^0_1$, the ${\tilde \tau}_1$, and possibly the
${\tilde \chi}^0_2$ and the ${\tilde \chi}^\pm_1$.
The best-fit CMSSM point is quite similar to the benchmark point
SPS~1a~\cite{sps}, which is close to point B of \citere{bench} and
has been shown to offer good
experimental prospects for both the LHC and ILC~\cite{lhcilc}.

The minimum values of $\chi^2$ are 2.5 for $\tb = 10$ and 2.8 for 
$\tb = 50$, found for $m_{1/2} \sim 320, 570 \gev$ and 
$A_0 = + m_{1/2}, - m_{1/2}$, respectively, revealing no
preference for either large or small $\tb$~
\footnote{
In our previous
analysis, we found a slight preference for $\tb = 10$ over $\tb = 50$.  
This preference has now been counterbalanced by the
increased pressure exerted by the Higgs mass constraint.
}%
. This also holds for intermediate $\tb$ values, see \citere{ehow4}
for details.

%%%%%%%%%%%%%%%%%%%%%%%%%%%%%%%%%%%%%%%%%%%%%%%%%%%%%%%%%%%%%%%%%%%%%
%%%%%%%%%%%%%%%%%%%%%%%%%%%%%%%%%%%%%%%%%%%%%%%%%%%%%%%%%%%%%%%%%%%%%

\section{Future evolution}
\label{sec:mtfuture}

\begin{figure}[htb!]
\begin{center}
\includegraphics[width=.48\textwidth]{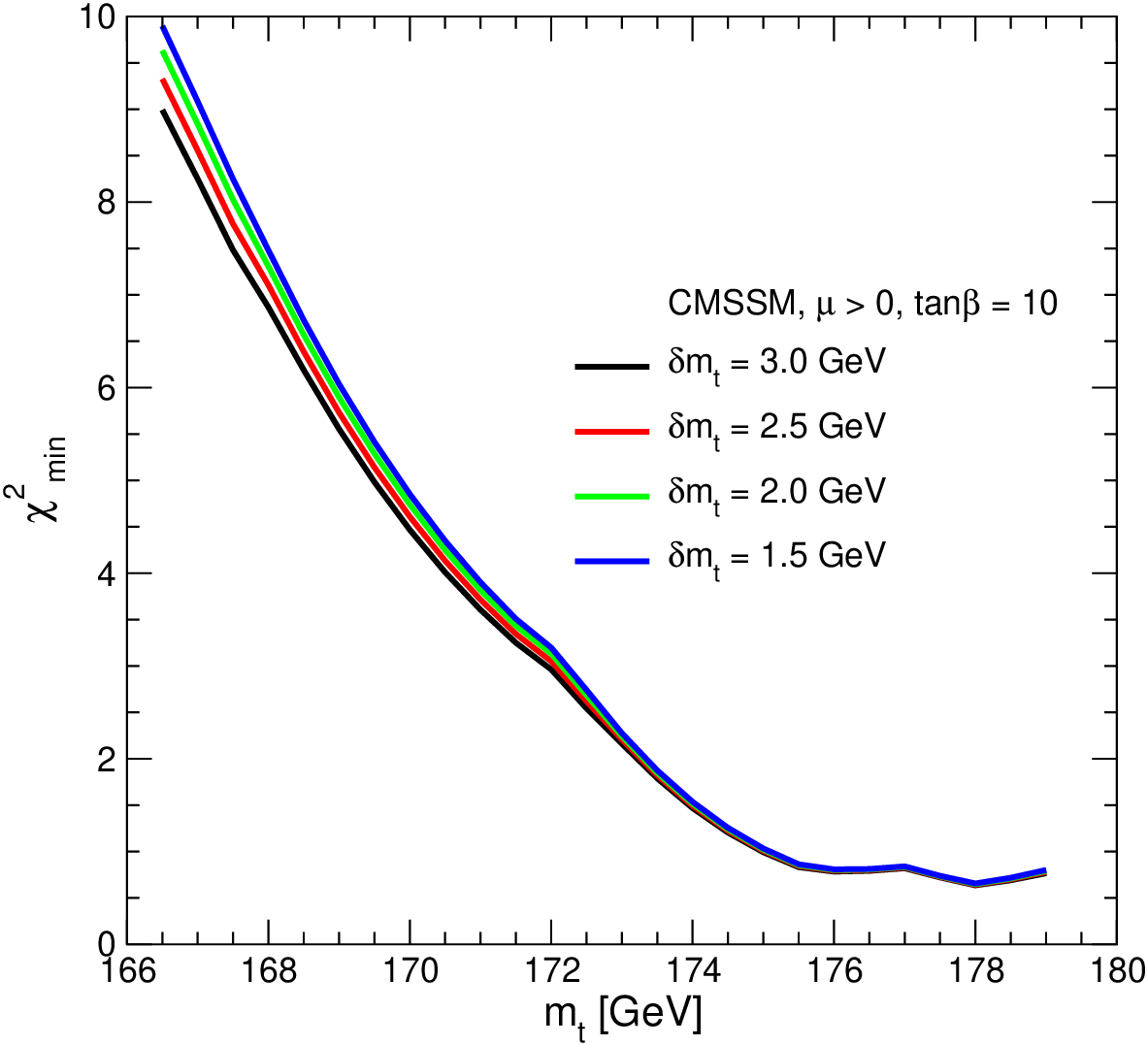}
\includegraphics[width=.48\textwidth]{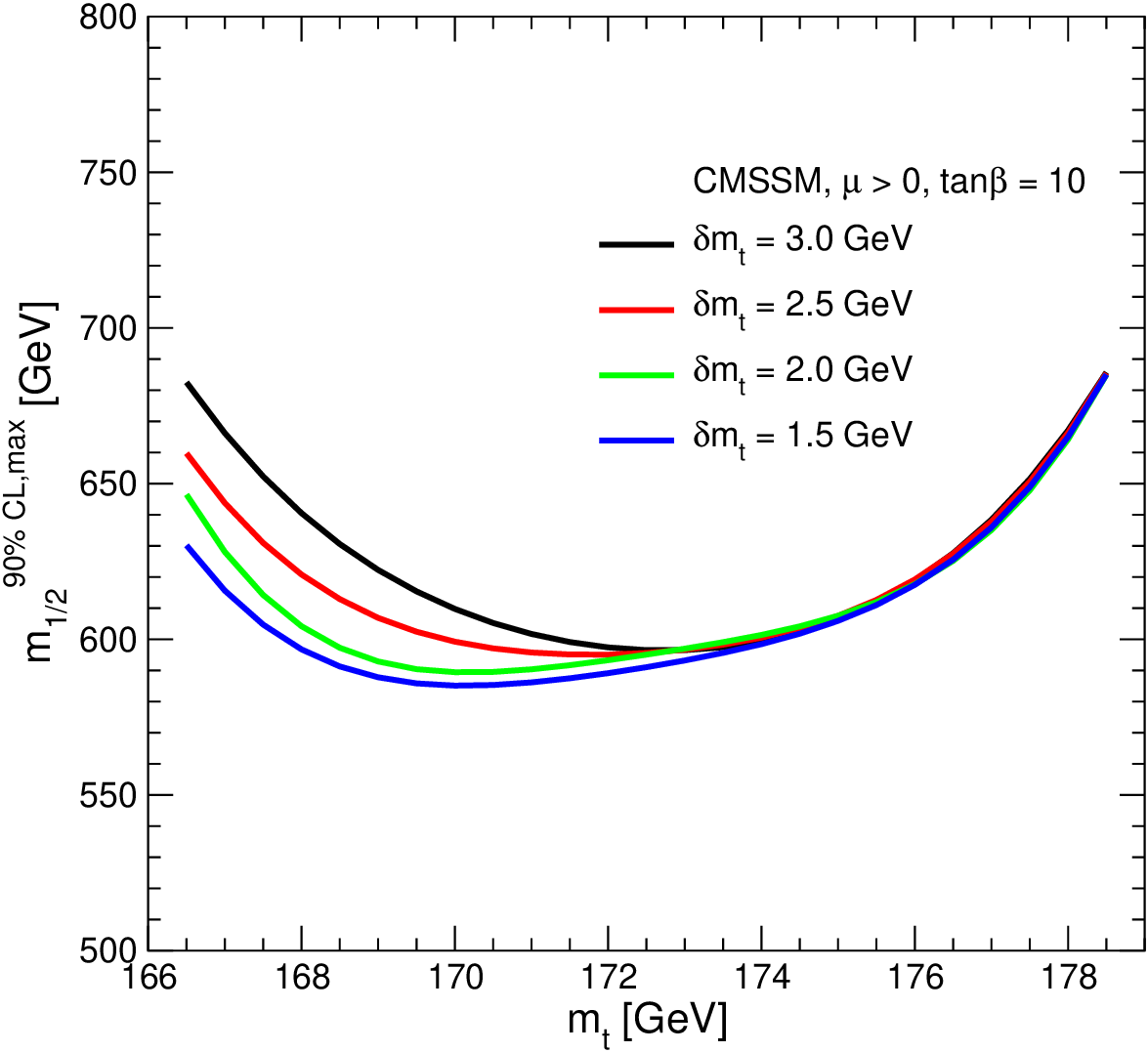}
\caption{
\it The dependence of (a) the minimum value of the $\chi^2$ distribution,
$\chi^2_{\rm min}$, and (b) the 90\% C.L.\ upper
limit for $m_{1/2}$ on $\mt$ and its experimental error $\delta \mt$, 
keeping the experimental values and theoretical predictions 
for the other precision observables
unchanged.
} 
\label{fig:varymt10} 
\end{center}  
\end{figure}

In view of the possible future evolution of both the central value of 
$\mt$ and its experimental uncertainty $\delta \mt$, we have analyzed the 
behaviour of
the global $\chi^2$ function for $166 \gev < \mt < 179 \gev$ and
$1.5 \gev < \delta \mt < 3.0 \gev$ for the case of $\tb = 10$ 
(assuming that the experimental results and theoretical predictions for 
the precision observables are otherwise unchanged),
as seen in the left panel of \reffi{fig:varymt10}. 
We see that the minimum value of $\chi^2$ is
almost independent of the uncertainty $\de\mt$, but increases
noticeably as the assumed central value of $\mt$ decreases. This effect
is not strong when $\mt$ decreases from $178.0 \gev$ to $172.7 \gev$,
but does become significant for $\mt < 170 \gev$. This effect is not
independent of the known preference of the ensemble of precision
electroweak data for $\mt \sim 175 \gev$ within the
SM~\cite{lepewwg,LEPEWWG}, to 
which the observables $\MW$ and $\sweff$ used here make important
contributions. On the other hand, as already commented, within the 
CMSSM there is the additional effect that the best fit values of $m_{1/2}$ 
for very low $\mt$ result in $\Mh$ values that are excluded by the LEP Higgs
searches~\cite{LEPHiggsSM,LEPHiggsMSSM} and have a very large
$\chi^2_{\Mh}$, resulting in an increase of the
lowest possible $\chi^2$~value for a given top-quark mass value. 
This effect also increases the value of  
$m_{1/2}$ where the $\chi^2$ function is minimized.
On the other hand, the right panel in \reffi{fig:varymt10} demonstrates that 
the 90\% C.L.\ upper
limit on $m_{1/2}$ shows only a small variation, less than $\sim 10\%$ for
$\mt$ in the preferred range above $170 \gev$~%
\footnote{
The plot has been
obtained by putting a smooth polynomial through the otherwise slightly
irregular points.
}%
. Finally, we note that the upper limit 
on $m_{1/2}$ is essentially independent of $\delta \mt$ for the
preferred range $\mt \gsim 170 \gev$. Thus for the latest experimental
value, $\mt = 172.5 \pm 2.3 \gev$ the results for the preferred
$m_{1/2}$ range remain essentially unchanged as compared to our analysis here.

%%%%%%%%%%%%%%%%%%%%%%%%% F I G U R E %%%%%%%%%%%%%%%%%%%%%%%%%%%%%%%%%%%%%%%%%
%%%%%%%%%%%%%%%%%%%%%%%%% F I G U R E %%%%%%%%%%%%%%%%%%%%%%%%%%%%%%%%%%%%%%%%%

It is striking that the preference noted earlier for relatively low values
of $m_{1/2}$ remains almost unaltered after the 
change in $\mt$ and the change in the
treatment of the LEP lower limit on $\Mh$. There seems to be little chance
at present of evading the preference for small $m_{1/2}$ hinted by the
present measurements of $\MW$, $\sweff$
%, $\br(b \to s \ga)$
and 
$(g - 2)_\mu$, at least within the CMSSM framework. It should be noted
that the preference for a relatively low SUSY scale is correlated with 
the top mass value lying in the interval 
$170 \gev \lsim \mt \lsim 180 \gev$.

%%%%%%%%%%%%%%%%%%%%%%%%%%%%%%%%%%%%%%%%%%%%%%%%%%%%%%%%%%%%%%%%%%%%%
%%%%%%%%%%%%%%%%%%%%%%%%%%%%%%%%%%%%%%%%%%%%%%%%%%%%%%%%%%%%%%%%%%%%%

\section{Conclusions}

Precision electroweak data and rare processes have some sensitivity to the loop
corrections that might be induced by supersymmetric particles. Present data 
exhibit some preference for a relatively low scale of soft supersymmetry 
breaking: $m_{1/2} \sim 300 \ldots 600 \gev$. This preference is
largely driven by  
$(g - 2)_\mu$, with some support from measurements of $\MW$ and $\sweff$. 
Here we have presented a re-evaluation in the light of new 
measurements of $m_t$ and $\MW$, and a more complete treatment of the
information provided by the bound from the LEP direct searches for the Higgs
boson. The preference for $m_{1/2} \sim 300 \ldots 600 \gev$ is
maintained in the CMSSM~%
\footnote{
A more complete discussion, also including models with non-universal
Higgs masses or gravitino dark matter, is given in~\cite{ehow4}.
}%
. 

The ranges of $m_{1/2}$ that are preferred would correspond to gluinos and
other sparticles being light enough to be produced readily at the LHC. Many
sparticles would also be observable at the ILC in the preferred CMSSM
parameter space.
In this respect the measurement of
$\MW$ is increasing in importance, particularly in the light of the recent
evolution of the preferred value of $m_t$. Future measurements of
$\MW$  and $\mt$ at the Tevatron will be particularly important in
this regard.

%%%%%%%%%%%%%%%%%%%%%%%%%%%%%%%%%%%%%%%%%%%%%%%%%%%%%%%%%%%%%%%%%%%%%
%%%%%%%%%%%%%%%%%%%%%%%%%%%%%%%%%%%%%%%%%%%%%%%%%%%%%%%%%%%%%%%%%%%%%

\subsection*{Acknowledgements}
S.H.\ and G.W.\ thank P.~Bechtle and K.~Desch for detailed
explanations on how to obtain $\chi^2$ values from the SM Higgs boson
searches at LEP. We thank A.~Read for providing the corresponding
$CL_s$ numbers. The work of S.H.\ was partially supported by CICYT
(grant FPA2004-02948) and DGIID-DGA (grant 2005-E24/2). The work of
K.A.O.\ was partially supported by DOE grant DE-FG02-94ER-40823.

%%%%%%%%%%%%%%%%%%%%%%%%%%%%%%%%%%%%%%%%%%%%%%%%%%%%%%%%%%%%%%%%%%%%%
%%%%%%%%%%%%%%%%%%%%%%%%%%%%%%%%%%%%%%%%%%%%%%%%%%%%%%%%%%%%%%%%%%%%%

%\newpage

%%%%%%%%%%%%%%%%%%%%%%%%%%%%%%%%%%%%%%%%%%%%%%%%%%%%%%%%%%%%%%%%%%%%%
%%%%%%%%%%%%%%%%%%%%%%%%%%%%%%%%%%%%%%%%%%%%%%%%%%%%%%%%%%%%%%%%%%%%%

\end{document}